\documentclass[iop, apj, linenumbers]{emulateapj}

\usepackage{threeparttable}
\usepackage{multirow}
\usepackage{times}
\usepackage{enumitem}
\usepackage{url} 
\usepackage{color}
\usepackage{amsmath,bm}
\usepackage{float}
\usepackage{graphicx}
\usepackage{adjustbox}
\usepackage{hyperref}
\hypersetup{
    colorlinks,
    citecolor=blue,
    urlcolor=blue,
    linkcolor=blue,
}

\shortauthors{Shun-Xuan He et al.}

\begin{document}

\title{Photometric-Metallicity and Distance Estimates for $\sim$70,000 RR Lyrae Stars from the Zwicky Transient Facility }

\author{Shun-Xuan He\altaffilmark{1,2}}
\author{Yang Huang\altaffilmark{1,2,8}}
\author{Xin-Yi Li\altaffilmark{3,8}}
\author{Hua-Wei Zhang\altaffilmark{4,5}}
\author{Gao-Chao Liu\altaffilmark{6,8}}
\author{Timothy C. Beers\altaffilmark{7}}
\author{Hong Wu\altaffilmark{1,2}}
\author{Zhou Fan\altaffilmark{1,2}}
\date{July 2024}

\shorttitle{Photometric Metallicities and Distances for 70,000 RR Lyrae Stars}

\altaffiltext{1}{National Astronomical Observatories, Chinese Academy of Sciences, Beijing 100101, People’s Republic of China;}
\altaffiltext{2}{University of Chinese Academy of Sciences, Beijing 100049,  People's Republic of China; huangyang@ucas.ac.cn}
\altaffiltext{3}{Department of Astronomy, College of Physics and Electronic Engineering, Qilu Normal University, Jinan 250200, People's Republic of China; lixinyi@qlnu.edu.cn}
\altaffiltext{4}{Department of Astronomy, Peking University, Beijing 100871, People's Republic of China}
\altaffiltext{5}{Kavli Institute for Astronomy and Astrophysics, Peking University, Bejing 100871, People's Republic of China}
\altaffiltext{6}{Center for Astronomy and Space Sciences, China Three Gorges University, Yichang 443002, People's Republic of China; gcliu@ctgu.edu.cn}
\altaffiltext{7}{Department of Physics and Astronomy and JINA Center for the Evolution of the Elements (JINA-CEE), University of Notre Dame, Notre Dame, IN 46556, USA}
\altaffiltext{8}{Corresponding authour: Yang Huang, Xin-Yi Li, Gao-Chao Liu}

\begin{abstract}

Utilizing Zwicky Transient Facility (ZTF) data and existing RR Lyrae stars (RRLs) catalogs, this study achieves the first calibration of the $P-\phi_{31}-R_{21}-\text{[Fe/H]}$ and $P-\phi_{31}-A_{2}-A_{1}-\text{[Fe/H]}$ relations in the ZTF photometric system for RRab and RRc stars. We also re-calibrate the period-absolute magnitude-metallicity (PMZ) and period-Wesenheit-metallicity (PWZ) relations in the ZTF $gri$-bands for RRab and RRc stars. Based on nearly 4100 stars with precise measurements of $P$, $\phi_{31}$, $A_{2}$, and $A_{1}$, and available spectroscopic-metallicity estimates, the photometric-metallicity relations exhibit strong internal consistency across different bands, supporting the use of a weighted averaging method for the final estimates. The photometric-metallicity estimates of globular clusters based on RR Lyrae members also show excellent agreement with high-resolution spectroscopic measurements, with typical scatter of 0.15\,dex for RRab stars and 0.14\,dex for RRc stars, respectively. Using hundreds of local RRLs with newly derived photometric metallicities and precise Gaia Data Release 3 parallaxes, we establish the PMZ and PWZ relations in multiple bands. Validation with globular cluster RR Lyrae members reveals typical distance errors of 3.1\% and 3.0\% for the PMZ relations, and 3.1\% and 2.6\% for the PWZ relations for RRab and RRc stars, respectively. Compared to PMZ relations, the PWZ relations are tighter and almost unbiased, making them the recommended choice for distance calculations. We present a catalog of 73,795 RRLs with precise photometric metallicities; over 95\% of them have accurate distance measurements. Compared to Gaia DR3, approximately 25,000 RRLs have precise photometric metallicities and distances derived for the first time.

\end{abstract}

\section{Introduction}

RR Lyrae stars (RRLs) are classical short-period (0.2\,d\rm-1\,d), low-mass ($<$1\,$M_\odot$), pulsating variable stars that reside at the intersection of the horizontal branch and the instability strip. They are typically categorized into three subtypes based on their pulsation properties: fundamental mode RRLs (RRab), first-overtone RRLs (RRc), and double-mode RRLs (RRd). Characterized by their old age ($\gtrsim$10\,Gyr) and metal-poor composition, most RRLs are widely distributed throughout the Galaxy, including the bulge (e.g., \citealt{Pietrukowicz2015,Olivares2024}), thick disk (e.g., \citealt{Mateu2018}), stellar halo (e.g., \citealt{Liu2022,Garcia2024}), and substructures (including streams and local overdensities; e.g., \citealt{Wang2022,Ye2024,Sun2024}), and are also observed in nearby galaxies (e.g., \citealt{Soszynski2016,Jacyszyn-Dobrzeniecka2017}). Their relatively bright luminosities and large amplitudes of distinctive light curves make them readily identifiable, and they can serve as excellent Population II chemical tracers. Furthermore, they are well-known standard candles, with a tight period-absolute magnitude-metallicity (PMZ) relation in the infrared/mid-infrared bands. They also exhibit specific absolute magnitude-metallicity relations in the visual bands, owing to their insensitivity to bolometric corrections within the RRLs temperature range (obvious in the $V$-band and gradually diminishing beyond the $R$-band; \citealt{bono2003pulsational,bhardwaj2022rr}). Thanks to these attributes, RRLs are invaluable for probing the chemical and kinematic properties of the Milky Way and its neighbors, offering insights into their formation and evolutionary history (e.g., \citealt{Iorio2019,Ablimit2022}).

With the advancement of numerous photometric surveys, such as the Sloan Digital Sky Survey (SDSS; \citealt{York2000}) and the Sloan Extension for Galactic Understanding and Exploration (SEGUE; \citealt{Yanny2009, Rockosi2022}), the Two Micron All Sky Survey (2MASS; \citealt{Skrutskie2006}), the Pan-STARRS1 survey (PS1; \citealt{sesar2017machine}), the Zwicky Transient Facility (ZTF; \citealt{masci2018zwicky}), the SkyMapper Southern Survey (SMSS; \citealt{Wolf2018}), the Optical Gravitational Lensing Experiment (OGLE; \citealt{soszynski2019over}), the All-Sky Automated Survey for SuperNovae (ASAS-SN; \citealt{jayasinghe2021asas,christy2023asas}), and the Gaia mission (\citealt{clementini2023gaia}), large numbers of RRLs across an extensive volume of the Galaxy have been discovered. At present, the Gaia third data release (DR3) has provided the largest and most reliable all-sky catalog of $\sim$270,000 
RRLs, identified over a 34 month period with multi-color data involving the $G$-, $BP$- and 
$RP$-bands (\citealt{clementini2023gaia}). However, this sample of RRLs remains incomplete; due to Gaia's scanning strategy, different sky regions have been monitored at varying epochs. In areas with sparse sampling, it becomes difficult to identify RRLs from their light curves, and even when identified, it is challenging to derive precise Fourier parameters from their curves, which are crucial for determining photometric-metallicity estimates. On the other hand, the ZTF, an optical sky survey characterized by extensive sky coverage, impressive depth, and a large number of observational epochs, is a treasure trove for the studies of variable stars (\citealt{bellm2017unblinking}). However, the rapid influx of data has resulted in a notable delay in the search and classification of variable stars. For RRLs, there are only a few classification studies based on the ZTF DR2 (\citealt{chen2020zwicky,Cheung2021}) and ZTF DR3 (\citealt{huang2022identifying}), despite the fact that ZTF data releases have advanced beyond DR20. There are likely to be numerous potential RRLs waiting to be discovered.

Metallicity is an important parameter for RRLs, and serves as a probe of the early chemical-evolution and formation history of our Galaxy. To derive [Fe/H] for RRLs, several spectroscopic methods have been developed. The most accurate method involves analyzing high-resolution spectra observed at certain phases (e.g., \citealt{sneden2011chemical,pancino2015chemical,gilligan2021metallicities}), but this is inefficient in terms of precious high-resolution spectroscopic observation time and resources. The $\Delta$S method, originally proposed by \citet{preston1959spectroscopic}, is applicable to low-to-moderate dispersion spectra (e.g., \citealt{Butler1975,Walker1991,Layden1994}), but it may lead to unexpected random and systematic errors in some cases where the $\Delta$S index and [Fe/H] have a nonlinear scale relation. 
\citet[][hereafter \hyperlink{Liu2020}{L20}]{liu2020probing} employed a novel template-matching method, deriving metallicities for more than 5000 RRLs down to [Fe/H] $\sim -3.0$, with a typical uncertainty of 0.2\,dex, based on extensive low-resolution spectroscopic surveys. Recently, \citet{Wang2024} employed similar approach, deriving metallicities for approximately 11,500 RR Lyrae stars with comparable uncertainty. However, due to limitations on depth and potential selection effects from spectroscopic surveys, these methods still only provide metallicity estimates for a small fraction of recognized RRLs. 

For the above reasons, the correlation between morphological light-curve characteristics and metallicity, discovered by \citet{clement1992rr,clement1993rr}, 
\citet{kovacs1995new}, and \citet{jurcsik1996determination}, has been explored and refined in many photometric systems through subsequent observational and theoretical work (e.g., \citealt{Smolec2005,nemec2013metal,ngeow2016palomar,Jurcsik2023}), and offers a more straightforward and cost-effective way to obtain [Fe/H] estimates. Recently, \citet[][hearafter \hyperlink{li2023photometric}{Li23}]{li2023photometric} calibrated a new period-Fourier parameters-[Fe/H] relation in the Gaia photometric system, deriving photometric metallicities for more than 130,000 RRLs using Fourier parameters from the Gaia RRL catalog with a typical uncertainty similar to that of low-resolution spectroscopic results. Recent studies employing machine learning or deep learning methods have also demonstrated improved performance in estimating 
[Fe/H] \citep[e.g.,][]{hajdu2018data,dekany2021metallicity,dekany2022photometric,Muraveva2024,monti2024leveraging}. 
Moreover, establishing a new relationship using the ZTF photometric system will allow us to obtain metallicity estimates for a larger number of RRLs, particularly as future ZTF data releases are expected to uncover numerous additional RRLs.

Since RRLs serve as excellent standard candles, extensive research has been conducted to calibrate the PMZ relation across various photometric systems. With the advent of large-scale infrared missions, the PMZ relations have been well established in the $JHK_{\rm s}$-bands and $W_1W_2$-bands,
achieving excellent precision, at the level of few percent (e.g., \citealt{  bono2003pulsational,catelan2004rr,muraveva2015new,muraveva2018rr,
bhardwaj2021rr,mullen2023rr}). For the optical bands, the relations are also well-established in various filters, such as the $BVRI$- and Gaia $G$-bands (e.g., \citealt{caputo2000pulsational,bono2003pulsational,braga2015distance,muraveva2015new,muraveva2018rr,bhardwaj2022rr,li2023photometric}). To further mitigate the impact of extinction on distance estimates from the PMZ relation, the Period-Wesenheit-Metallicity (PWZ) relation was developed. It builds upon multi-band PMZ calibrations by incorporating the Wesenheit magnitude—an extinction-free parameter constructed from multi-band magnitude combinations (e.g., \citealt{Madore1982,Madore1991}). This refinement provides an extinction-independent calibration, enabling more precise distance measurements, particularly in high-extinction regions. Similar to the PMZ relations, the PWZ relation has been rigorously calibrated across various photometric systems, spanning optical ($BVRI$, Gaia $G$, $G_{BP}$, $G_{RP}$) to infrared ($JHK_{\rm s}$, $W_1W_2$) bands, achieving comparable precision levels (e.g., \citealt{Neeley2019,Garofalo2022,mullen2023rr,Zgirski2023}). However, calibration efforts for both the PMZ and PWZ relations using Sloan or Sloan-like filters remain limited, with only a few studies (e.g., \citealt{sesar2017machine,vivas2017absolute,bhardwaj2021optical}) focusing on specific globular clusters, or relying on small samples of globular clusters (GCs) that require further refinement. Recently, \citet{ngeow2022zwicky} calibrated the PMZ relation using 755 RRLs from 46 globular clusters observed with the ZTF, and derived a PWZ relation from the same dataset. However, such relations, based on member stars of globular clusters, may face several issues: 1) photometric blending due to crowded fields; 2) the presence of multiple populations of varying metallicity; and 3) a limited number of calibrators, leading to sparse parameter coverage. Therefore, similar calibrations with field RRLs is important and necessary.

In this study, we have collected photometric data from ZTF for over 70,000 RRLs. From this sample, approximately 4,100 RRLs with precise metallicity estimates were selected based on the updated catalog of \hyperlink{Liu2020}{L20}. By combining these metallicities with the derived ZTF parameters ($P$, $\phi_{31}$, $A_{2}$, and $A_{1}$), we have calibrated the $P-\phi_{31}-R_{21}-\text{[Fe/H]}$ relations for RRab stars and $P-\phi_{31}-A_{2}-A_{1}-\text{[Fe/H]}$ relations for RRc stars in the ZTF $gri$-bands, respectively. Then, using several hundred local RRLs with photometric-metallicity from the new relations and accurate distances from Gaia parallax measurements, we have further re-determined the PMZ and PWZ relations in the three bands for RRab and RRc stars, respectively. 

The structure of this paper is as follows. In Section \ref{sec:samp}, we introduce the preprocessing of photometric data and the selection process for our samples. In Section \ref{sec:cali}, we detail the calibration of the $P-\phi_{31}-R_{21}-\text{[Fe/H]}$ and $P-\phi_{31}-A_{2}-A_{1}-\text{[Fe/H]}$ relations, as well as the PMZ and PWZ relations in the $gri$-bands for RRLs, and provide photometric-metallicity and distance estimates for the entire sample. Additionally, this section includes internal comparisons across different bands and comparisons with other studies. In Section \ref{sec:vali}, we provide multiple external validations for the derived photometric-metallicity and distances. Section \ref{sec:cat} presents the final RRL catalog with descriptions. Finally, Section \ref{sec:summ} provides a summary of this work.

\section{Data and Sample}\label{sec:samp}
In this section, we construct two primary sample sets for this work. The first set comprises RRLs from the ZTF photometric data. We use a portion of this set to calibrate the metallicity and distance relations, which are then applied to the entire set to determine the parameters for each star. The second set includes RRLs with metallicity estimates derived from spectroscopic surveys; these stars are used to calibrate the photometric metalliciy relations. Both sample sets undergo a well-defined selection process before use, as described below.

\subsection{The ZTF RRL Sample}

\subsubsection{Data Sources and Verification}
To obtain a larger pool of usable sample stars, we integrate preliminary RRL candidates using RR Lyrae catalogs from multiple photometric surveys as follows:

\begin{itemize}[leftmargin=*]

\item The Gaia DR3 RRLs table: This table ontains 270,905 RRLs from Gaia observational data beginning in 2013. They are classified into subtypes `RRab', `RRc', and `RRd' using multi-band ($G$-, $BP$-, and $RP$- band) data and the Specific Objects Study (SOS) Cep\&RRL pipeline (\citealt{clementini2023gaia}). Given the widespread use of these sample, we employ them as a primary source for our work.

\item The ZTF DR3 RRLs table: This table contains 71,755 RRLs classified by the Fourier parameters and random forest method (\citealt{huang2022identifying}) from the ZTF data and reported as part of the Gaia early data release 3 (EDR3). This catalog has high completeness and purity relative to the Gaia DR2 RRLs (\citealt{clementini2019gaia}), and achieves more accurate period determinations than the ZTF DR2 RRL sample (\citealt{chen2020zwicky}), with 96\% of the sample stars in common, thanks to the increased number of observational epochs. Since subtypes of RRL stars are not provided, we only select stars with a probability score more than 0.8 (max is 1.0) and sub-classify them visually to ensure their purity.

\item The ASAS-SN RRLs table: This catalog is combined from two tables from ASAS-SN data in the $V$- (\citealt{jayasinghe2021asas}) and $g$- (\citealt{christy2023asas}) bands, respectively. The $V$-band table contains 45,060 RRL candidates with $V \lesssim$ 17; a newer $g$-band table contains 26,515 RRL candidates with $g \lesssim$ 18.5. The two tables share stars in common, but the $g$-band sample stars generally performs with higher confidence, due to the higher cadence of $g$-band data. As before, we only select those stars with probability score more than 0.8 for both `RRab' and `RRc' type stars.

\item The PS1 RRLs table: This table contains 239,044 RRLs classified by template-fitting techniques and a machine learning algorithm (\citealt{sesar2017machine}) from the sparse, asynchronous multi-band PS1 3$\pi$ survey, which probes deeper than ZTF, with similar sky coverage. In this table, the subtypes for each sample are provided with `RRab' and `RRc' final classification scores. Note that sum of the two scores is not equal to 1.0 because of RRd stars participating in the training process.

\end{itemize}

We combine these tables to pre-select RRLs. Given the critical role of period determination and the re-calculation method described in Section \ref{sec:p_recal}, we adopt the catalog-reported periods as initial values. For sources appearing in multiple catalogs, we prioritize the most reliable period based on catalog credibility, following this descending hierarchy: Gaia, ZTF, ASAS-SN, and PS1. This ranking reflects the relative accuracy of period measurements, which depends primarily on photometric precision and sampling cadence. Gaia is assigned the highest priority due to its superior photometric precision and moderate sampling cadence ($\sim$40 epochs on average). ZTF and ASAS-SN follow, as they provide an order-of-magnitude higher sampling while maintaining sufficient photometric quality. PS1 is given the lowest priority due to its sparse observational cadence ($\lesssim$12 epochs over 4.5 years). Moreover, visual inspection of the light curves indicates that this low cadence not only reduces the reliability of period determination, but also results in a high contamination rate from eclipsing binaries (EBs).  
To mitigate this contamination, we remove potential EBs from the combined table using EB catalogs from Gaia, ZTF, and OGLE (\citealt{mowlavi2023gaia,chen2020zwicky,soszynski2016ebs}), finding that 37\% of PS1 RR Lyrae candidates are actually EBs.  
In the following sections, we apply stricter selection criteria to further exclude other contaminants, ultimately constructing a high-purity, high-quality sample.

Finally, considering the photometric quality of the ZTF data, we have only downloaded the light curves in the $gri$-bands for those candidates that have more than 50 high-quality photometric observation epochs (catflags$=$0) (\citealt{masci2018zwicky}).

\subsubsection{Period Re-calculation}\label{sec:p_recal}
The period is a critical feature in light-curve analysis. However, the accuracy of period determination is influenced by the sampling cadence, and for ground-based surveys like ZTF and ASAS-SN, aliased periods may also occur due to the limited observational window (e.g., as discussed in \citealt{chen2020zwicky}). To minimize the presence of aliased periods and ensure consistency in periods derived from the $gri$-bands, we fine-tune the initial period values within 0.001 days using the Generalized Lomb-Scargle algorithm (GLS; \citealt{lomb1976least,scargle1982studies,Ferraz-Mello1981,Zechmeister2009}). Subsequently, we select sample stars with correct periods based on the constraints of false-alarm probability and manual inspection of phase-folded light curves. For sample stars with ambiguous light variations, we expand the period search range to the typical theoretical range of RRLs: 0.2 to 1 day for type RRab stars, and 0.2 to 0.6 day for type RRc stars, and then once again employ the GLS to identify the correct period. After this round of screening and elimination, we exclude a portion of the non-RRLs, retaining only those stars with accurate periods.

In order to minimize contamination, all light curves in our sample have undergone manual inspection, and candidates primarily consisting of EBs and RRd have been excluded. Due to the absence or unreliability of classification scores in ZTF DR3 and PS1, we manually classified these sample stars into RRab and RRc types. These sample stars undergo a more refined selection and labeling in Section \ref{sec:FFL}.

\subsubsection{Fourier Decomposition, Filtering, and Labeling} \label{sec:FFL}
Fourier decomposition parameters can parametrically represent the morphological light-curve characteristics, playing a crucial role in obtaining the physical parameters of variable stars, particularly for RRLs. Using the decomposition equation:
\begin{equation}
    m(t) = m_{0} + \sum_{i=1}^{k} A_{i} \cos\left(2\pi i t / P + \phi_{i}\right)\text{,}
\end{equation} we employ the known periods ($P$) and the \texttt{lcfit} package by \citet{istvandekany_2022}, followed by a Gaussian Process Regression (GPR) of the phase-folded light curves, to derive the key parameters for the three bands, including mean magnitude ($m_0$), peak-to-peak amplitude ($A_{\rm tot}$), amplitude ratios ($R_{i1} = A_{i} / A_{1}$), phase differences ($\phi_{i1} = \phi_{i} - i\phi_{1}$), and their errors. Furthermore, phase coverage (Phcov) and signal-to-noise ratio (SNR) are also output as indicators to evaluate the distribution and fitting quality of the photometric points in the light curve. Additionally, we introduce the standard deviation of residuals (Rstdev) as a new metric. Detailed descriptions of these indicators are presented in Table\,\ref{tbl:Fouriercat}.

Based on the parameters obtained, we further filtered the data to yield a high-quality sample set. Finally, we obtained a total of 73,796 sample stars (52,572 type RRab and 21,224 type RRc stars; hereafter the ZTF\_RRL\_ALL sample) after applying the following cuts: i) Phocv $>$ 0.9; ii) SNR $>$ 50; iii) 0.2\,d$< P <$1\,d for RRab or 0.2\,d$< P <$ 0.6\,d for RRc stars. The distribution diagrams of the Fourier parameters in the $g$-band with best period (i.e., the mean of period in $gri$-bands; referred to as `period’ below if the band is not specified) are shown in Figure\,\ref{fig:Fourier_g}; similar distributions for the $r$- and $i$-bands are shown in Appendix Figure\,\ref{fig:Fourier_r} and Figure\,\ref{fig:Fourier_i}. We then derive the physical parameters of these sample stars.

\begin{figure*}[!htbp]
\centering
\includegraphics[width=0.98\linewidth]{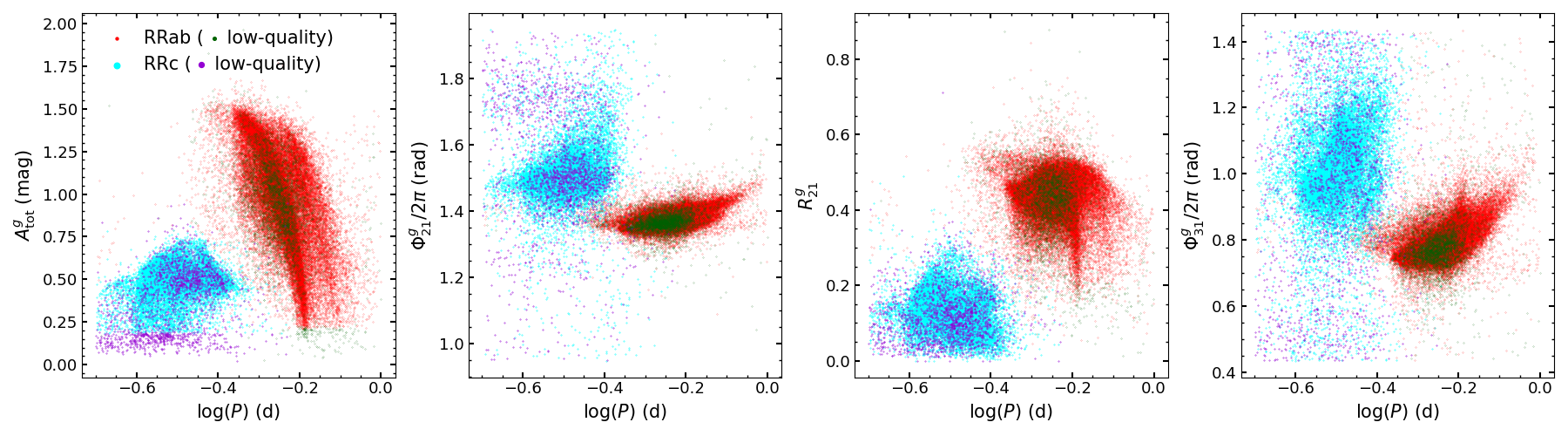}
\caption{Distribution of the Fourier parameters ($A_{\rm tot}$, $\phi_{21}$, $R_{21}$, $\phi_{31}$), as a function of the period in the $g$-band, for 73,796 ZTF\_RRL\_ALL sample stars (52,572 type RRab and 21,224 type RRc stars). In each panel, type RRab stars and their relatively low-quality sample stars are indicated in red and dark green, and type RRc stars and their relatively low-quality sample stars are indicated in cyan and dark purple. Figures for the $r$- and $i$-bands are presented in Appendix \ref{fig:Fourier_r} and \ref{fig:Fourier_i}, respectively.}
\label{fig:Fourier_g}
\end{figure*}

Furthermore, we labeled relatively low-quality sample stars with potentially problematic light curves in some bands, partially following the method discussed in \citet{ngeow2022zwicky}. In detail, they introduced several flags to label low-quality RRL stars based on light curves. For example, the 'A' flag designates outliers in the distribution of amplitudes across the $gri$-bands, as well as amplitude ratios in band pairs; the 'C' flag marks outliers in the period-color and period-extinction-free \textit{Q}-index ($Q=(\bar{m}_g-\bar{m}_r)-\mathrm{R}(\bar{m}_r-\bar{m}_i)$; $\bar{m}_{g,r,i}$ represents the mean magnitude of sample stars) relations. Following their method, we first labeled outlier sample stars with 'A' flags based on minimum-maximum amplitude cuts in the $gri$-bands (cut values referenced from \citealt{ngeow2022zwicky}, Figure\,3) and the iterative 3$\sigma$ clipping of amplitude ratios across different band pairs. Then, we labeled outliers with 'C' flags solely based on the period-$Q$ relation, where the extinction coefficient provided by \citet{wang2019optical} was used to calculate $\mathrm{R}=$1.443. We have introduced a new 'D' flag to identify excessive dispersion in the residuals of fitting light curves, as indicated by the Rstdev value. Sample stars are labeled with 'D' flags when Rstdev exceeds 0.15/0.10 in the $g$-band and 0.12/0.08 in the $r$-band for type RRab and RRc stars, respectively. Finally, we marked 6645 (4420 type RRab, 2225 type RRc) outlier sample stars that have one or more flags. Because these stars may be affected by blending and other photometric problems, they are excluded for calibrating 
photometric-metallicity and absolute-magnitude relations. The remaining sample is referred to as the ZTF\_RRL\_FIT sample in the following. The Fourier parameters catalog for ZTF\_RRL\_ALL sample is accessible on Zenodo\footnote{\href{https://doi.org/10.5281/zenodo.14561442}{10.5281/zenodo.14561442}} and a description of the columns is shown in in Table\,\ref{tbl:Fouriercat}. Additionally, we provide a master table (Table\,\ref{tbl:sample_def}) summarizing all samples and subsamples in this work, including their abbreviations and descriptions, for consolidated reference.

\begin{table}
\centering
\caption{Overview of Samples and Subsamples: Abbreviations and Description\label{tbl:sample_def}}
\begin{threeparttable}
\begingroup 
\setlength{\tabcolsep}{5mm} 
\renewcommand{\arraystretch}{1.5} 
\begin{tabular}{lp{4.5cm}}
\hline 
Abbreviation & Description \\ 
\hline
ZTF\_RRL\_ALL & All sample stars \\
ZTF\_RRL\_FIT & Sample stars used for the subsequent calibration process\\
ZTF\_RRL\_FIT\_META & Calibration sample stars used for the photometric-metallicity relation \\
ZTF\_RRL\_FIT\_DIS1 & Calibration sample stars used for the PMZ relation\\
ZTF\_RRL\_FIT\_DIS2 & Calibration sample stars used for the PWZ relation \\
\hline
\end{tabular}
\endgroup 
\end{threeparttable}
\end{table}

\subsection{Spectroscopy-based Metallicity Sample}
To calibrate the $P-$Fourier parameters$-$[Fe/H] relations in the $gri$-bands (Section \ref{sec:cali_meatl}), we select the sample by cross-matching the ZTF\_RRL\_FIT sample with the updated spectroscopic sample of 10,916 RRLs from \hyperlink{Liu2020}{L20}, using the following criteria:
\begin{itemize}[leftmargin=*]

\item The spectral signal-to-noise ratios of the sample must be greater than 20. This cut balances the need to retain a sufficient number of stars for calibration while ensuring high-quality spectra for precise [Fe/H] estimates.

\item The measurement errors of [Fe/H] for the sample must be smaller than 0.15\,dex for both RRab and RRc stars.

\item The difference in period between our results and the reference period from \hyperlink{Liu2020}{L20} must be smaller than 0.001 day.
 
\item The measurement error of $\phi_{31}$ must be smaller than 0.5\,rad, and the measurement error of $R_{21}$ must be smaller than 0.06 for both type RRab and RRc stars, except for the i-band of RRc. Given the limited RRc sample in the $i$-band, the measurement errors of $R_{21}$ is set to 0.1.

\end{itemize}

Finally, we select 2875 RRab stars and 1182 RRc stars for the calibration, hereafter referred to as the ZTF\_RRL\_FIT\_META sample. The initial numbers of sample stars to be fitted in different bands are shown as N$_{\text{ini}}$ in Table\,\ref{tbl:fitcoff_metal}.

\vspace{1\baselineskip}

\section{Photometric Estimation of Metallicity and Distance}\label{sec:cali}
In this section, we calibrate the period-Fourier parameters-metallicity ([Fe/H]) relations in the 
$gri$-bands for RRab and RRc stars using the ZTF\_RRL\_FIT\_META sample described previously. Accurate measurements of period and Fourier parameters, particularly $R_{21}$ and $\phi_{31}$, enable the derivation of metallicities for stars in the ZTF\_RRL\_ALL sample. Furthermore, employing local RRLs with precise distances from Gaia parallax measurements and the photometric metallicities determined here, we construct the PMZ and PWZ relations for RRLs in the $gri$-bands. These relations facilitate the calculation of absolute magnitude and distance for the entire sample of stars with photometric-metallicity estimates.

\subsection{Metallicity}\label{sec:cali_meatl}

In the ongoing development of photometric-metallicity estimators, our work builds upon 
\citet{dekany2021metallicity}, who extensively explored the significance of Fourier parameters and their combined effects. This foundation has enabled us to establish an estimator for photometric-metallicity using ZTF data.

Based on \citet{dekany2021metallicity}, and our follow-up tests, we construct linear models for RRab and RRc stars across the $gri$-bands. For RRab stars, the models are parameterized by $P$, $R_{21}$, and $\phi_{31}$. In the case of RRc stars, $R_{21}$ is further separated into $A_2$ and $A_1$. The models are expressed as follows:
\begin{equation}
{\rm[Fe/H]} =a_{0} + a_{1}P + a_{2}\phi_{31} + a_{3}R_{21}\text{,}
\end{equation}
\begin{equation}
{\rm[Fe/H]} =a_{0} + a_{1}P + a_{2}\phi_{31} + a_{3}A_{2} + a_{4}A_{1}\text{,}
\end{equation}
where [Fe/H] is the spectroscopic metallicity estimated by \hyperlink{Liu2020}{L20}, and $a_{i}$ ($i$ = 0, ..., 4) are the fitting coefficients. Note that, although \citet{dekany2021metallicity} recommended using $A_{2}$ for RRab stars, we have chosen to use $R_{21}$ instead. This decision was based on $R_{21}$'s superior performance in the actual fitting process. For RRc stars, the parameters used are in agreement with expectations from \citet{dekany2021metallicity}.

\begin{figure*}[!htbp]
\centering
\includegraphics[width=0.98\linewidth]{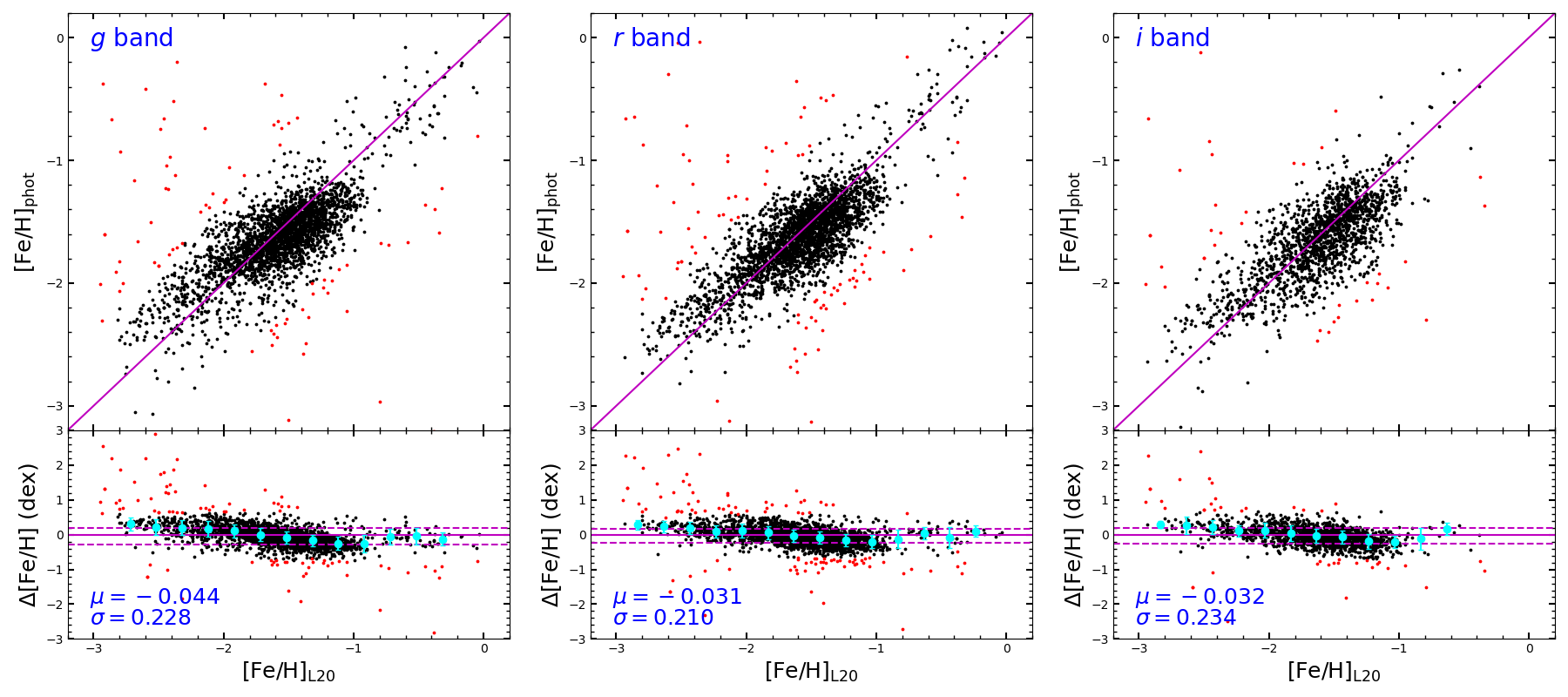}
\caption{The top section of the panels compare metallicity estimates for RRab stars between L20 and our photometric estimators, with results for the $gri$-bands shown from left to right, respectively. Red dots in each panel represent sample stars excluded by the 3$\sigma$ clipping, while black dots are those included in the final fitting. The solid-purple line is the one-to-one line. The lower section of each panel shows $\Delta$[Fe/H] differences (our work minus L20), with the mean and standard deviation displayed at the bottom left of each panel. Purple-dashed lines denote the 1$\sigma$ boundaries, and the solid-purple line indicates the mean residual level. The cyan dots and associated error bars represent the median $\Delta$[Fe/H] values and their standard deviations for each [Fe/H]$_{\mathrm{L20}}$ bin.}
\label{fig:metal_cali_ab}
\end{figure*}

\begin{figure*}[!htbp]
\centering
\includegraphics[width=0.98\linewidth]{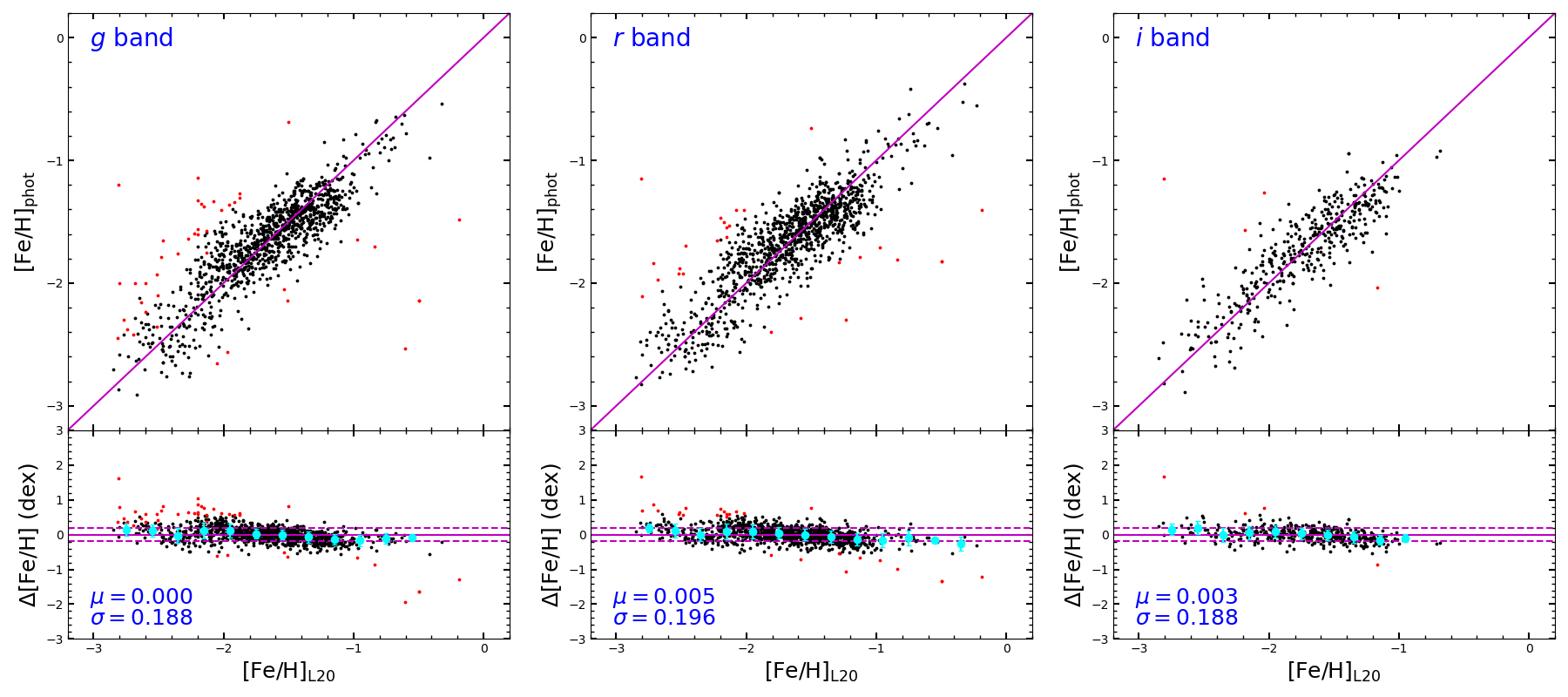}
\caption{Similar to Figure\,\ref{fig:metal_cali_ab}, but for RRc stars.}
\label{fig:metal_cali_c}
\end{figure*}

The metallicity distribution in the ZTF\_RRL\_FIT\_META sample is non-uniform, with a primary concentration between [Fe/H] $\approx -2.0$ to $-1.0$, a tail extending toward lower metallicity, and a smaller, distinct group at the metal-rich end, as shown in Figures \ref{fig:metal_cali_ab} and \ref{fig:metal_cali_c}. The clustering and metal-poor tail correspond to the well-known Oosterhoff dichotomy of RR Lyrae stars \citep{Oosterhoff1939,Arp1955,Preston1959}, likely arising from a complex interplay between stellar and Galactic evolution \citep[e.g.,][]{Zhangshan2023}. Meanwhile, the rare metal-rich RRLs ([Fe/H] $> -1.0$) remain an open question, with hypotheses suggesting origins via binary evolution \citep[e.g.,][]{Karczmarek2017,Bobrick2024} or accretion from nearby galaxies \citep[e.g.,][]{Feuillet2022}. To minimize the impact of this uneven 
sampling distribution on the fitting process, we employed a weighted fitting method based on the sample's density, as suggested by the works of \citet{dekany2021metallicity} and 
\citet{li2023photometric}. Briefly, the weight ($\omega_{d}$) for each star is determined by the normalized Gaussian kernel density ($\rho_{d}$) derived from the metallicity distribution. An empirical density threshold value, $\rho_{d0}$, divides the sample into two parts: stars from the portion with density greater than $\rho_{d0}$ are assigned $\omega_{d} = 1/\rho_{d}$, while the other portion retains a constant value $\omega_{d} = 1/\rho_{d0}$. The value of $\rho_{d0}$ is set to 0.2 for both RRab and RRc stars. The fitted coefficients for the $gri$-bands are presented in Table\,\ref{tbl:fitcoff_metal}, following an iterative 3$\sigma$ clipping and weighted fitting procedure. A comparison of metallicity estimates from \hyperlink{Liu2020}{L20} and our photometric estimator is shown in Figure\,\ref{fig:metal_cali_ab} (RRab) and Figure\,\ref{fig:metal_cali_c} (RRc), with the $gri$-band results displayed from left to right.

The bias and scatter of the fitting results are listed in Table\,\ref{tbl:fitcoff_metal}. For RRab stars, the estimator exhibits a minor bias of 0.03$-$0.04\,dex and a scatter of 0.21$-$0.23\,dex across different bands, which is comparable to \hyperlink{li2023photometric}{Li23}'s findings of 0.24\,dex. In the case of RRc stars, there is a negligible bias with comparable scatter of 0.19$-$0.20\,dex, again in line with \hyperlink{li2023photometric}{Li23}'s result of 0.19\,dex. However, upon binning the sample, a mild bias is evident at the metal-poor end and around [Fe/H] = $-$1, particularly for RRab stars, despite the substantial weights allocated to these regions. This bias is a common challenge in the calibration of unevenly sampled data, and is also observed in similar studies (e.g., \citealt{dekany2021metallicity,dekany2022photometric,li2023photometric}). An increased number of metal-rich and extremely metal-poor stars in future studies is expected to reduce biases in the calibration caused by sample diversity.

Using the established relations, we calculated the photometric metallicities for the ZTF\_RRL\_ALL sample across the $gri$-bands. We conducted internal comparisons of [Fe/H] values from the three bands, as illustrated in Appendix Figure\,\ref{fig:metal_comself_ab} (RRab) and Figure\,\ref{fig:metal_comself_c} (RRc). Given the slight deviations across bands, we utilized a weighted average approach to ascertain the metallicity ([Fe/H]\_weig\_TW), with weights determined by the inverse square of the uncertainty in metal abundance for each band. This uncertainty stems from both random errors in the Fourier parameters and fitting-coefficient errors, estimated via 1000 Monte Carlo (MC) simulations, as well as methodological errors across bands. Ultimately, we estimated metallicities for 73,795 RRLs (52,571 RRab and 21,224 RRc stars), which is only one star less than the ZTF\_RRL\_ALL sample, due to the exclusion of abnormal $R_{21}$ values equal to zero.

\begin{table*}
  \centering
  \scriptsize
  \caption{Fitted Coefficients for Photometric-metallicity Relations\label{tbl:fitcoff_metal}}
  \begin{tabular*}{\textwidth}{@{\hspace{2mm}}@{\extracolsep{\fill}}ccccccccc@{\hspace{2mm}}}
    \hline
    \hline
    Band & 
    $a_{0}$ & 
    $a_{1}$ & 
    $a_{2}$ & 
    $a_{3}$ & 
    $a_{4}$ & 
    $\mu$ & 
    $\sigma$ & 
    N$_{\mathrm{fit}}/$N$_{\mathrm{ini}}$ \\
    \hline
    \multicolumn{9}{c}{RRab} \\
    $g$   & $-6.33\pm0.03$ & $-5.00\pm0.02$   &  $1.37\pm0.01$  & $1.79\pm0.03$  & \dots  &  $-0.044$  &  0.228  & 2681/2768 \\
    $r$   & $-5.39\pm0.03$ & $-5.99\pm0.02$  &  $1.22\pm0.00$  & $1.65\pm0.03$  & \dots  &  $-0.031$  &  0.210  & 2738/2844 \\
    $i$   & $-4.73\pm0.03$ & $-6.18\pm0.03$  &  $1.06\pm0.01$  & $1.39\pm0.04$  & \dots  &  $-0.032$  &  0.234  & 1581/1624 \\
    \multicolumn{9}{c}{RRc} \\
    $g$   & $-0.93\pm0.10$  & $-9.59\pm0.11$  &  $0.30\pm0.01$  & $-13.66\pm0.59$  & $3.74\pm0.23$ &  0.000  &  0.188  & 1051/1094 \\
    $r$   & $-1.12\pm0.10$  & $-9.66\pm0.11$  &  $0.36\pm0.01$  & $-14.94\pm0.77$  & $3.81\pm0.34$ &  0.005  &  0.196  & 1070/1097 \\
    $i$   & $0.53\pm0.19$  & $-9.85\pm0.19$  &  $0.17\pm0.02$  & $-23.88\pm1.37$  & $3.18\pm0.79$ &  $0.003$  &  0.188  & 407/411 \\
    \hline
  \end{tabular*}
  
  \vspace{0.5mm}
  \parbox{\textwidth}{
    \scriptsize
    \tablecomments{The fitting parameters $a_{i}$ ($i = 0, \dots, 4$) are defined in equations (1) and (2) for RRab and RRc stars. 
    $\mu$ and $\sigma$ represent the bias and dispersion of the fitted relation, respectively. 
    N$_{\mathrm{ini}}$ and N$_{\mathrm{fit}}$ denote the number of sample stars before and after the $3\sigma$-rejection algorithm.}
  }
\end{table*}

\subsection{Distance}
Utilizing precise distances from Gaia parallax measurements and photometric-metallicity estimates from this study, we calibrate the PMZ relation in the $gri$-bands using local RRLs. Additionally, we calibrate the PWZ relations, accounting for the significant effects of extinction, as described in Subsection \ref{sec:cali_pwz}.

\subsubsection{The PMZ Relations}\label{sec:cali_pmz}
We select the calibration sample by cross-matching the ZTF\_RRL\_FIT sample with Gaia DR3 data, based on the following selection criteria:

\begin{itemize}[leftmargin=*]

\item The RRLs must exhibit parallaxes greater than 0.2 mas (about 5\,kpc), and have relative parallax errors less than 10\%

\item The RRLs, whether of type RRab or RRc, must be located in high-latitude regions with $|b| \geq 25^{\circ}$, and the value of $E(B-V)$ must be less than 0.1\,mag according to 
\citet[][hereafter \hyperlink{schlegel1998maps}{SFD98}]{schlegel1998maps}

\item The errors of the photometric metallicities from the $r$-band ([Fe/H]\_r\_err\_TW) must be less than 0.25\,dex and 0.29\,dex for RRab and RRc stars, respectively

\end{itemize}

The first two criteria ensure accurate determination of absolute magnitudes in the $gri$-bands by accounting for the precision of the distance and extinction measurements. The third criterion ensures a reliable photometric-metallicity estimate. Notably, we use $r$-band data for the metallicity-error criterion, intentionally forgoing a multi-band weighted average. This approach ensures that lower-quality sample stars are excluded, while higher-quality ones, even if represented in fewer bands, are retained, as the multi-band weighted average error is generally smaller than the single-band error. Finally, we select 343 RRab and 173 RRc stars for the calibration exercise, hereafter referred to as the ZTF\_RRL\_FIT\_DIS1 sample.

\begin{figure*}[!htbp]
\centering
\includegraphics[width=0.98\linewidth]{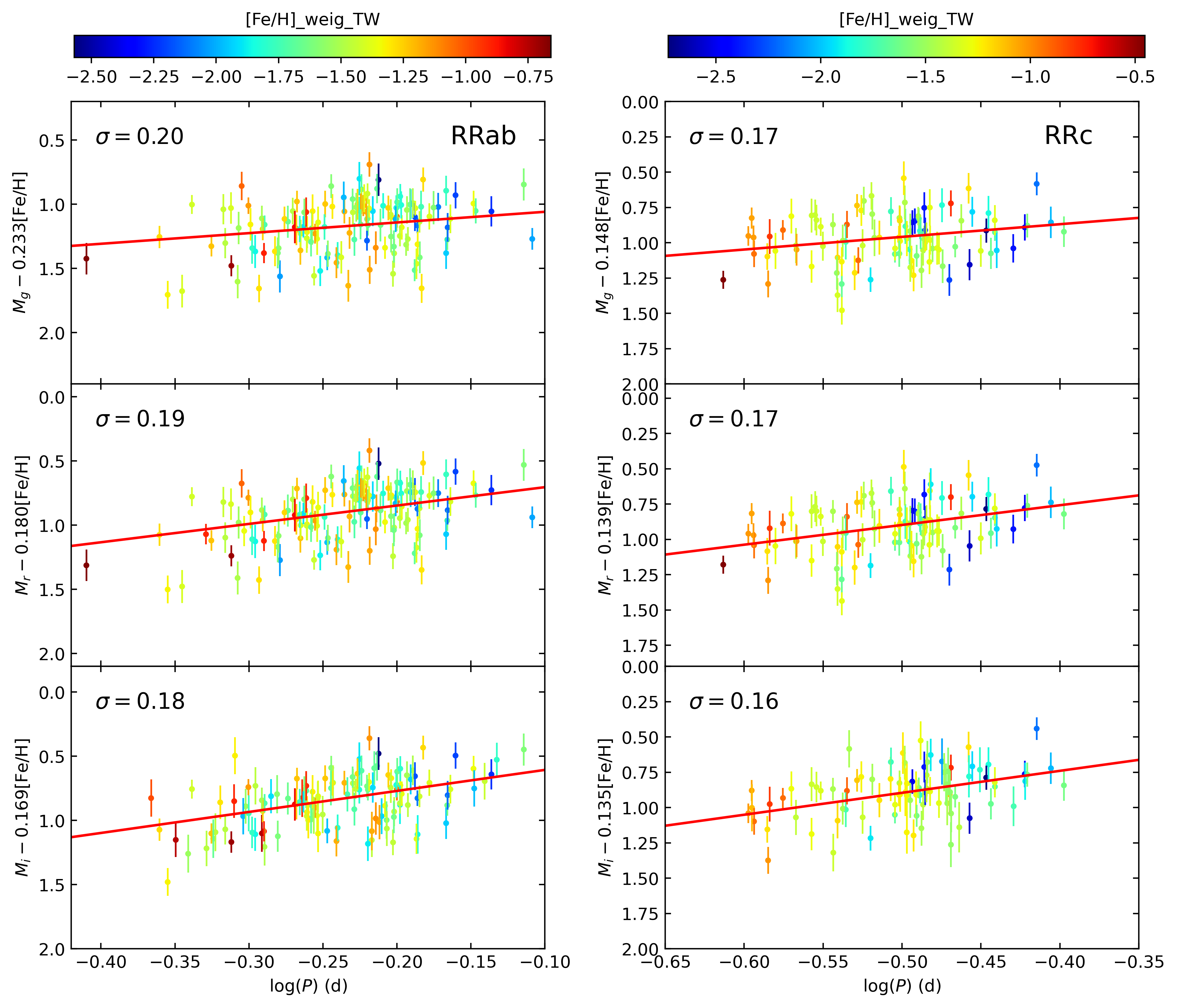}
\caption{The PMZ relations for the ZTF\_RRL\_FIT\_DIS1 sample, refined by absolute-magnitude error constraints, are presented in two columns of panels, color-coded by metallicity. The left and right columns correspond to RRab and RRc types, respectively, displaying the $gri$-bands from top to bottom. The solid-red line in each panel denotes the best fit, and the standard deviation of the residuals is provided in the top left corner or each panel.}
\label{fig:pmz_2dfit}
\end{figure*}

Prior to carrying out the calibration, we estimate the distances to these sample stars using the corrected Gaia DR3 parallaxes (\citealt{lindegren2021gaia}). These distances were derived employing the Bayesian method provided by \citet{huang2021discovery}. Subsequently, the absolute magnitudes in the $gri$-bands for the ZTF\_RRL\_FIT\_DIS1 sample were calculated using these distances, the mean apparent magnitudes ($m_{0}$), and the extinction corrections from \hyperlink{schlegel1998maps}{SFD98}. To ensure sample quality, we set stringent error limits on the absolute magnitudes for both RRab and RRc stars. Specifically, we restricted the errors, which include contributions from distance and mean magnitude, to be less than 0.13 for the $g$- and $r$-bands. For the $i$-band, the error limits were set to 0.15 for RRab stars and 0.18 for RRc stars. The remaining sample was then used to calibrate the PMZ linear relation:
\begin{equation}
M_{g,r,i} = a\,{\rm log}(P) + b\,{\rm[Fe/H]}+ c\text{,}
\end{equation}
where $log(P)$ is the logarithm of the pulsation period calculated as described above, and [Fe/H] is the photometric-metallicity estimate determined in Section \ref{sec:cali_meatl}. The resulting fit coefficients $a$, $b$, $c$, and the number of sample stars used for calibration N$_{\text{fit}}$ in the $gri$-bands are listed in Table\,\ref{tbl:fitcoff_PMZPWZ}. The fits are shown in Figure\,\ref{fig:pmz_2dfit}, with the left panel for type RRab stars, and the right panel for type RRc stars.

\begin{table*}
\scriptsize
\centering
\caption{Fitted Coefficients for the PMZ and PWZ Relations}
\label{tbl:fitcoff_PMZPWZ}
\begin{tabular*}{\textwidth}{@{\hspace{2mm}}@{\extracolsep{\fill}}cccccc@{\hspace{2mm}}}
  \hline
  \hline
  & $a$ & $b$ & $c$ & $\sigma$ & N$_{\mathrm{fit}}$ \\
  \hline
  \multicolumn{6}{c}{RRab} \\
  $M_{g}$   & $-0.829\pm0.171$ &  $0.233\pm0.030$  &  $0.976\pm0.071$  &  0.197  & 141 \\
  $M_{r}$   & $-1.432\pm0.169$ &  $0.180\pm0.029$  &  $0.561\pm0.071$  &  0.189  & 145 \\
  $M_{i}$   & $-1.639\pm0.182$ &  $0.169\pm0.032$  &  $0.442\pm0.076$  &  0.177  & 128 \\
  $W_{gr}$  & $-2.678\pm0.133$ &  $0.007\pm0.018$  &  $-0.610\pm0.055$  &  0.215  & 136 \\
  $W_{gi}$  & $-2.671\pm0.171$ &  $0.056\pm0.027$  &  $-0.266\pm0.072$  &  0.183  & 147 \\
  $W_{ri}$  & $-2.512\pm0.170$ &  $0.095\pm0.027$  &  $-0.006\pm0.073$  &  0.190  & 152 \\
  \multicolumn{6}{c}{RRc} \\
  $M_{g}$   & $-0.898\pm0.254$ &  $0.148\pm0.028$  &  $0.509\pm0.159$  &  0.170  & 99 \\
  $M_{r}$   & $-1.395\pm0.254$ &  $0.139\pm0.028$  &  $0.202\pm0.159$  &  0.167  & 100 \\
  $M_{i}$   & $-1.551\pm0.276$ &  $0.135\pm0.032$  &  $0.120\pm0.172$ &  0.162  & 88  \\
  $W_{gr}$  & $-3.437\pm0.237$ &  $0.019\pm0.037$  &  $-1.209\pm0.158$  &  0.221  & 103 \\
  $W_{gi}$  & $-3.186\pm0.285$ &  $0.107\pm0.040$  &  $-0.740\pm0.186$  &  0.172  & 96 \\
  $W_{ri}$  & $-3.136\pm0.281$ &  $0.137\pm0.040$  &  $-0.544\pm0.184$ &  0.189  & 98  \\
  \hline
\end{tabular*}
\vspace{0.5mm}
\parbox{\textwidth}{
    \scriptsize
    \tablecomments{The fitting parameters $a$, $b$, $c$ are defined in equation (3) and (4), while $\sigma$ represents the dispersion of the fitted relation. The quantity N$_{\mathrm{fit}}$ denotes the number of sample stars used for calibration in different bands.}
}
\end{table*}

Overall, the fits depicted in Figure\,\ref{fig:pmz_2dfit} align with expectations based on prior observations and theoretical works (e.g., \citealt{muraveva2015new,braga2015distance,sesar2017machine}). Specifically, as the bandpass shifts towards redder wavelengths, the fit reveals an increased dependency on the period term (indicated by a larger $\lvert a \rvert$), a decreased reliance on metallicity (reflected in a smaller $b$), and a tighter PMZ relationship (indicated by a smaller $\sigma$). Moreover, the RRc relation in the same band is tighter than that for RRab stars, suggesting that distance calibration relationships using RRc stars may be more precise, as discussed by \citet{li2023photometric}. Further detailed validation concerning relative distance errors is explored in Section \ref{sec:vali}.

We also compare our results with published PMZ relations in the $gri$-bands, noting that there is relatively less research in the ZTF bands compared to others. The main previous research for comparison includes studies by \citet[][hereafter \hyperlink{sesar2017machine}{S17}]{sesar2017machine} and \citet[][hereafter \hyperlink{ngeow2022zwicky}{N22}]{ngeow2022zwicky}. \hyperlink{sesar2017machine}{S17} calibrated the relations using 55 RRab stars from five GCs in the PS1 photometric system, which is also used for ZTF data. \hyperlink{ngeow2022zwicky}{N22} calibrated relations for both RRab and RRc stars, utilizing ZTF data similar to ours, focusing on 755 RRL GC members with metallicities ranging from [Fe/H] = $-$2.4 to $-$0.9. Building on these comparisons, we further analyze the dependencies on the period and metallicity terms.

\begin{figure*}[!htbp]
\centering
\includegraphics[width=0.98\linewidth]{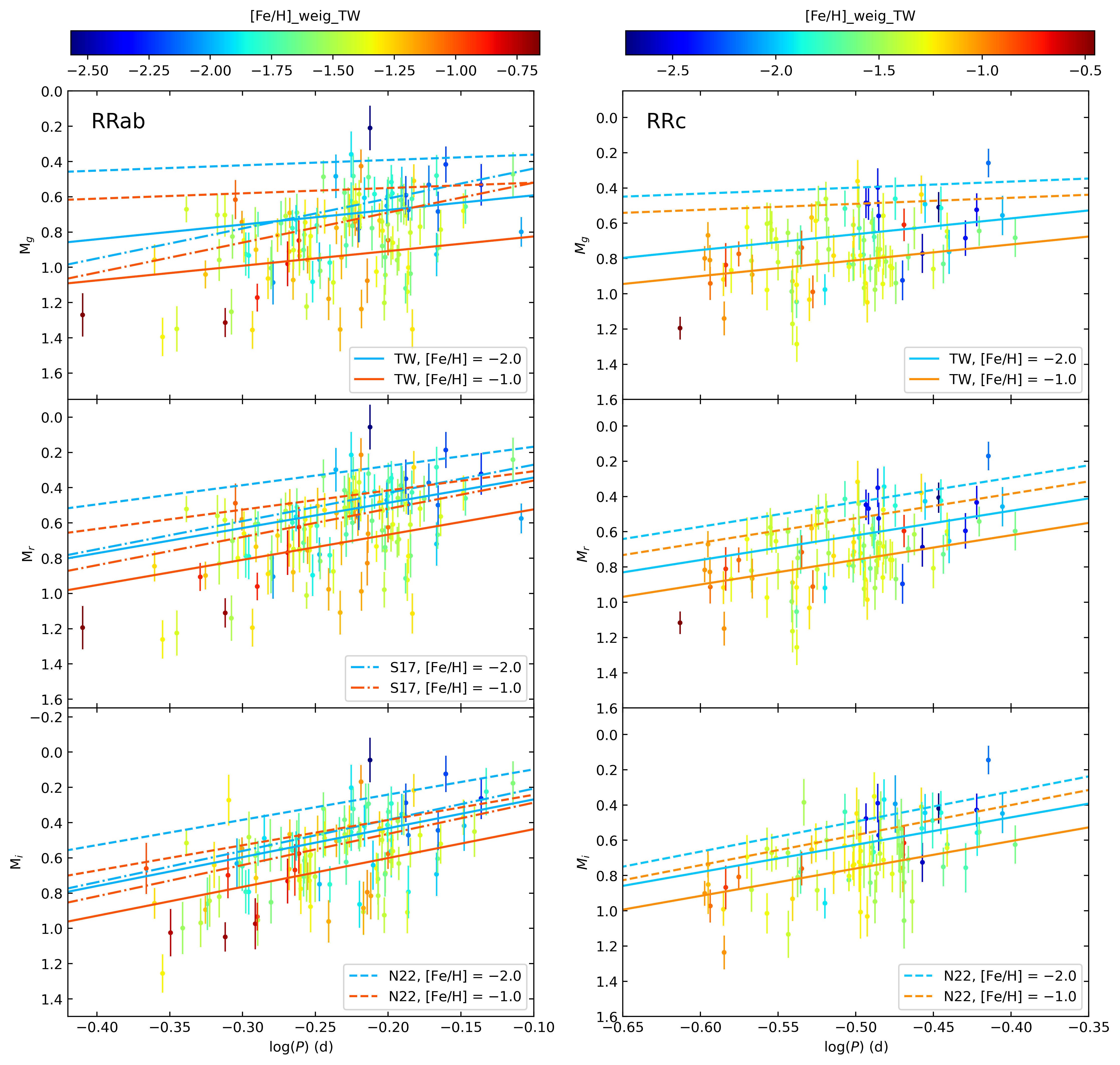}
\caption{The PMZ relation, similar to Figure\,\ref{fig:pmz_2dfit}, but with an adjusted y-axis to display only absolute magnitude without the metallicity term, showing the sample distributed in the Period-Absolute Magnitude space. The fitted PMZ relations are evaluated at [Fe/H] = $-$2.0 and
$-$1.0, and depicted in two distinct colors. For comparison, the relations from this work (TW), S17, N22, at these fixed [Fe/H] values are represented by the solid, dash-dot, and dashed lines, respectively. The slopes of the lines indicate the dependence on the period terms, while the vertical separation between two lines from the different [Fe/H] level reflects the metallicity term's influence.}
\label{fig:pmz_ref}
\end{figure*}

To facilitate a clear comparison of slopes of the period term, we present the fitted PMZ relations at [Fe/H] = $-$2.0 and $-$1.0 in Figure\,\ref{fig:pmz_ref}. A distinct slope disagreement is evident in the $g$-band between our work and \hyperlink{sesar2017machine}{S17} ($\sim$2.5$\sigma$ for RRab), and \hyperlink{ngeow2022zwicky}{N22} ($\sim$2$\sigma$ and $\sim$1.4$\sigma$ for RRab and RRc stars, respectively), with our coefficient $a$ falling midway between them. \hyperlink{ngeow2022zwicky}{N22} discussed the reason for the disagreement between their work and \hyperlink{sesar2017machine}{S17}; in the case of a relatively consistent performance in the $r$- and $i$-bands, they suggest that the most likely reason is the Bayesian inference method used by \hyperlink{sesar2017machine}{S17}, which employs a uniform prior for the $g$-band period term rather than a Gaussian prior for the $r$- and 
$i$-bands. 

Better confirmation comes from the agreement in the slope of the period term between \hyperlink{sesar2017machine}{S17} and our work in the $r$- and $i$-bands, indicating a possible susceptibility of the slope to the choice of priors. This may explain the difference in the slope of the period term for the $g$-band among the three approaches. With the exception of the 
$g$-band, the slopes of the period term exhibit better agreement with \hyperlink{sesar2017machine}{S17} for RRab stars, and are consistent with \hyperlink{ngeow2022zwicky}{N22} for RRc stars, despite the lack of calibration from \hyperlink{sesar2017machine}{S17}. Furthermore, the dependence of the metallicity term is discernible in the PMZ relations at fixed periods, as manifested by the vertical separation between lines representing different [Fe/H] levels in Figure\,\ref{fig:pmz_ref}. Generally, our relation shows a milder metallicity dependence than \hyperlink{sesar2017machine}{S17}, which has a weak dependence, and is slightly larger than \hyperlink{ngeow2022zwicky}{N22} ($\sim$1.3$\sigma$ for RRab stars and $\sim$1.4$\sigma$ for RRc stars, on average), possibly due to the difference of the metallicity distribution.

Based on the newly constructed PMZ relations, we estimate the distances for the entire sample. For high-latitude regions with $|b|\geq 25^{\circ}$, we adopt the \hyperlink{schlegel1998maps}{SFD98} map for extinction correction. To maximize the sample of available distance parameters, particularly for RRab stars within regions where $|b|< 25^{\circ}$, we utilize the converted absorption (A$_G$) provided by \citet[][hereafter \hyperlink{clementini2023gaia}{C23}]{clementini2023gaia}. These values are adjusted from the Gaia $G$-band to the $gri$-bands according to the extinction law provided by \citet{wang2019optical}. We then derived the distances for a subset of the ZTF\_RRL\_ALL sample with extinction values in the three bands. The uncertainty of the distances arises from two components: one is the random errors calculated through 1000 MC simulations, accounting for the uncertainty of the fit coefficients, the errors of photometric-metallicity, a fixed uncertainty of 0.05\,mag for $E(B-V)$, and the absorption errors provided by \hyperlink{clementini2023gaia}{C23} for some type RRab stars; the other is the methodological errors, reflected in the scatter of the PMZ relation across different bands. 

One sample may yield multiple values from the relations across different bands. Consequently, we conducted internal comparisons, similar to those for metallicity calculations, as illustrated in Appendix Figures \ref{fig:pmz_dist_comself_ab} and \ref{fig:pmz_dist_comself_c} for RRab and RRc stars, respectively. Notably, Figure\,\ref{fig:pmz_dist_comself_ab}, particularly the $d_g$ vs $d_r$ panel, reveals local higher dispersion. Our analysis suggests that this small but significant subset is attributed to the use of A$_G$ from \hyperlink{clementini2023gaia}{C23}. Given that these sample stars are characterized by high extinction values (A$_G>$ 1\,mag), they may be associated with greater errors. Additionally, for both RRab and RRc stars, the $r$-band distances consistently exhibit a larger offset relative to other bands. Despite this, the overall sample's acceptable agreement leads us to adopt a weighted average method, analogous to that for metallicity, where weights are determined by the inverse square of the distance uncertainty per band. Ultimately, we estimate distances for 56,593 RRLs in the entire sample: 49,505 RRab and 7088 RRc stars. This final sample size is significantly different from the ZTF\_RRL\_ALL sample, particularly for RRc stars, due to the limited availability of extinction values.

\vspace{1\baselineskip} 

\subsubsection{The PWZ Relations}\label{sec:cali_pwz}
Estimates of distance from PMZ relations are often affected by extinction, making it challenging to achieve precise measurements for samples with significant extinction-related deviations. In an attempt to enhance precision, we utilize the formally extinction-free Wesenheit magnitude to construct the PWZ relation, focusing on sample stars with photometric data in at least two bands. In this work, the Wesenheit magnitude are defined as:

\begin{eqnarray}
  W_{gr} & = & r - 2.712 (g-r) \nonumber \\
  W_{gi} & = & g - 2.193 (g-i) \nonumber \\
  W_{ri} & = & r - 3.914 (r-i), \nonumber 
\end{eqnarray}
 where the relative coefficients calculated adopted the Galactic extinction law derived by 
 \citet{wang2019optical}.

For the calibration of the PWZ relation, we construct our sample by cross-matching the ZTF\_RRL\_FIT sample with Gaia DR3 data, akin to our approach for the PMZ relations, and apply the following selection criteria:

\begin{itemize}[leftmargin=*]

\item The RRLs must contain photometric data in at least two bands

\item The RRLs must exhibit parallaxes greater than 0.2 mas (about 5\,kpc), and have relative parallax errors less than 10\%

\item The errors of the photometric metallicities from the $r$-band ([Fe/H]\_r\_err\_TW) must be less than 0.23\,dex and 0.27\,dex for RRab and RRc stars, respectively

\end{itemize}

The first criterion ensures the accessibility of the Wesenheit magnitude, while the second and third criteria are designed to enhance the precision of the measurements. The noteworthy effect of the third criterion is discussed further in Section \ref{sec:cali_pmz}. Here, we adopt smaller limit values, due to the relatively large numbers of sample stars that meet the remaining criteria. Finally, we selected 719 RRab and 349 RRc stars for the calibration, hereafter referred to as the ZTF\_RRL\_FIT\_DIS2 sample.

\begin{figure*}[!htbp]
\centering
\includegraphics[width=0.98\linewidth]{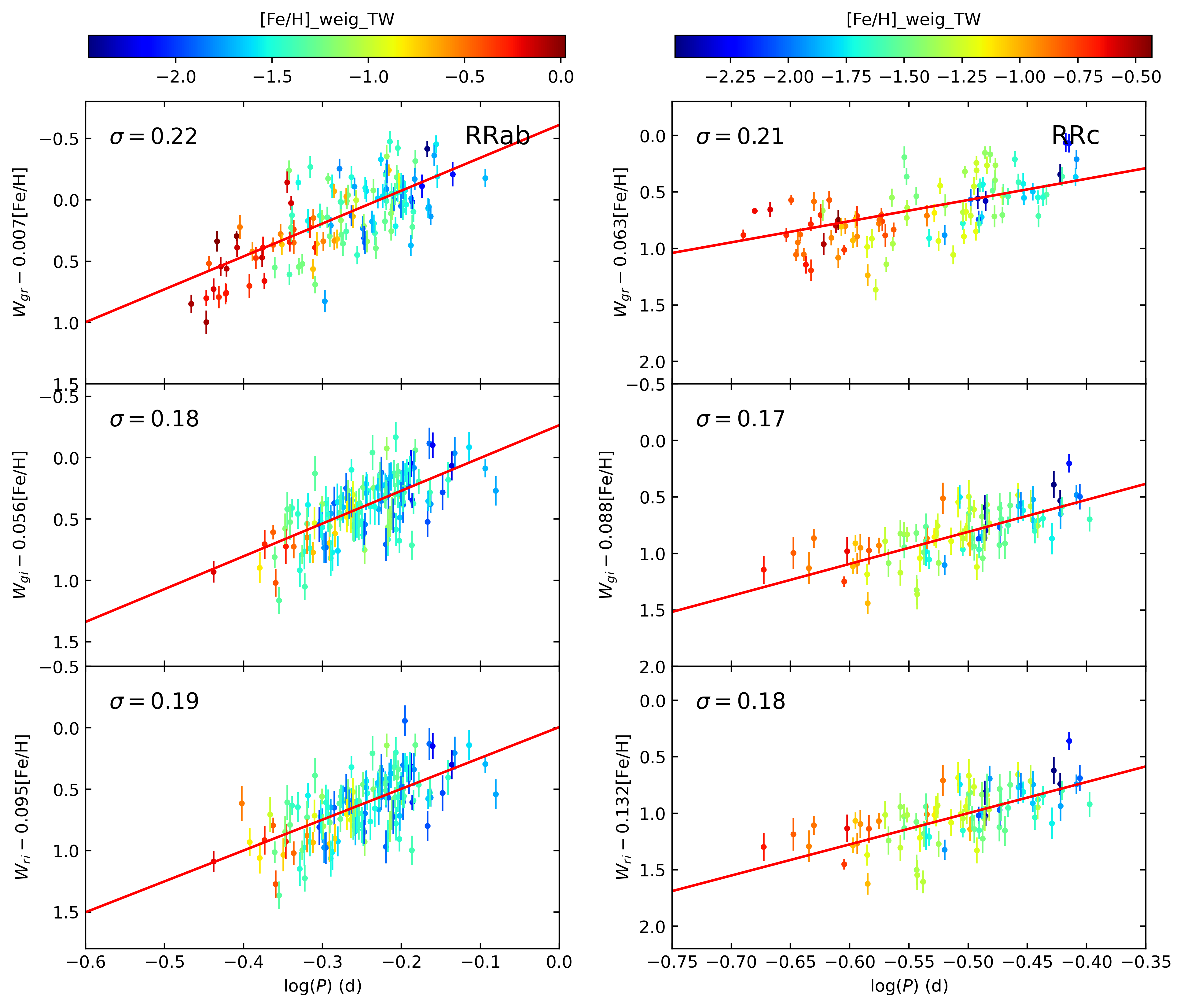}
\caption{The PWZ relations for the ZTF\_RRL\_FIT\_DIS2 sample, refined by absolute-magnitude error constraints, are presented in two columns of panels, color-coded by metallicity. The left and right columns correspond to RRab and RRc stars, respectively, displaying the $gr$-, $gi$-, and $ri$-band pairs from top to bottom. The solid-red line in each panel denotes the best fit, and the standard deviation of the residuals is shown in the top left corner of each panel.}
\label{fig:pwz_2dfit}
\end{figure*}

We determined the absolute Wesenheit magnitudes for these sample stars using the distances derived from the Bayesian method (\citealt{huang2021discovery}) and the mean apparent magnitudes in two bands, as mentioned previously. To ensure the reliability of our results, we imposed error limits on the absolute magnitudes, which account for uncertainties in distance and mean magnitude measurements. Specifically, we required these errors to be less than 0.10 mag for the $g$-band, and 0.15 mag for the $r$- and $i$-bands, for both RRab and RRc stars. After applying these criteria, we selected the remaining sample stars to calibrate the PWZ linear relation:
\begin{equation}
W_{gr,gi,ri} = a\,{\rm log}(P) + b\,{\rm[Fe/H]}+ c\text{,}
\end{equation}
where the variables are defined as in Section \ref{sec:cali_pmz}. The resulting fit coefficients $a$, $b$, $c$, and the number of sample stars used for calibration N$_{\text{fit}}$ in the $gri$-bands are listed in Table\,\ref{tbl:fitcoff_PMZPWZ}. The fits are shown in Figure\,\ref{fig:pwz_2dfit}; the left column of panels is for type RRab stars, while the right column of panels if for type RRc stars.

A notably shallow slope for RRc stars is evident, particularly in the $gr$-band pair, as shown in Figure\,\ref{fig:pwz_2dfit}. The primary reason for this phenomenon is that the $W_{gr}$ of the metal-rich ([Fe/H] $>-$1) sample stars is unexpectedly smaller than anticipated. Upon examining the light curves in the $g$- and $r$-bands, and the distance calculation process, we suspect that these metal-rich RRc sample may exhibit different PW$_{gr}$Z relationships. However, separate calibration for this subset is not robust due to the limited sample size and relatively low precision of the metallicity measurements.

\begin{figure*}[!htbp]
\centering
\includegraphics[width=0.98\linewidth]{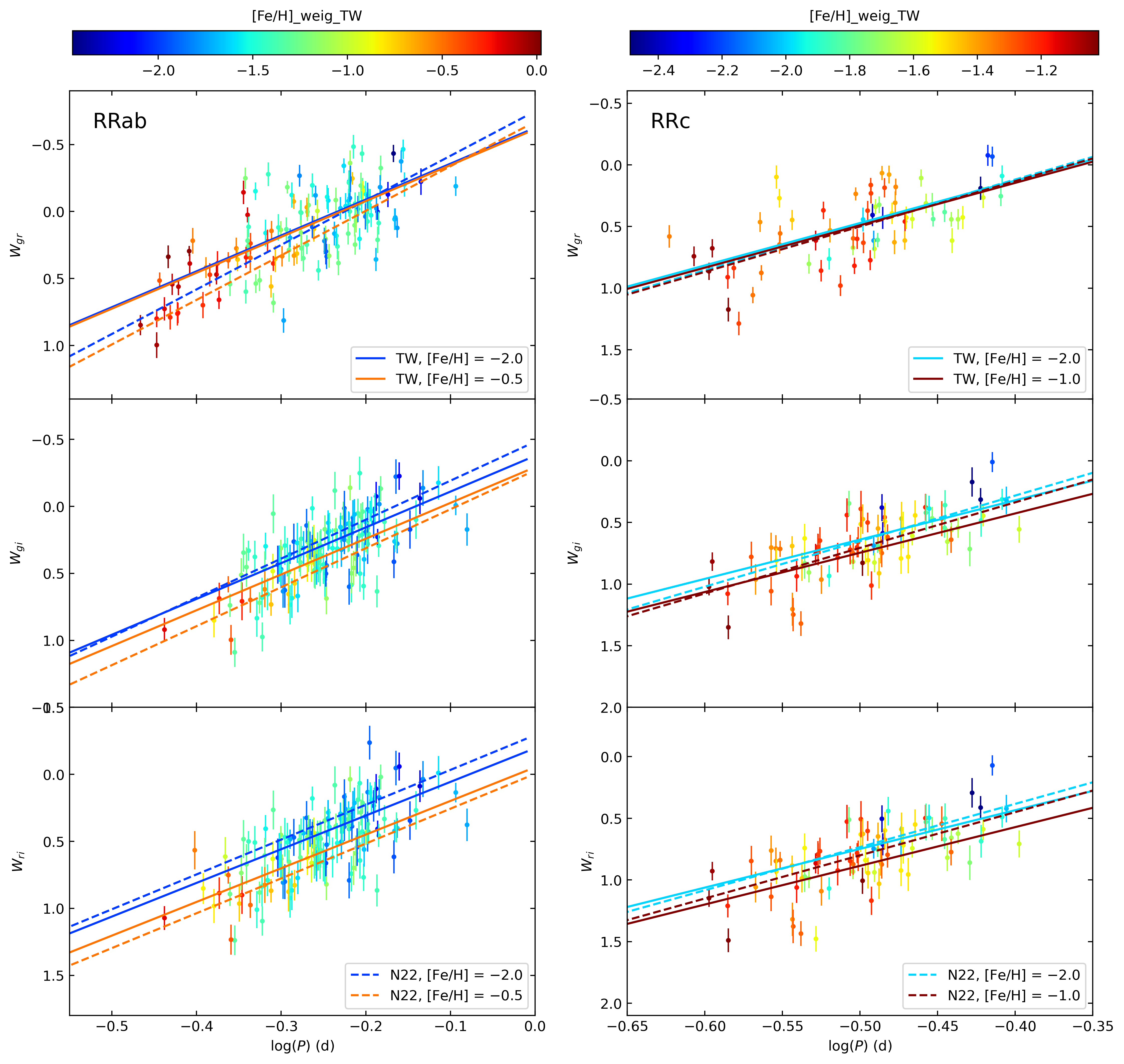}
\caption{The PWZ relations, similar to Figure\,\ref{fig:pwz_2dfit}, but with an adjusted y-axis to display only the absolute magnitude without the metallicity term, showing the sample distributed in the Period-Absolute Magnitude space. The fitted PWZ relations are evaluated at [Fe/H] = $-$2.0 and $-$0.5($-$1) for RRab and RRc stars, and depicted in two distinct colors. For comparison, relations from this work (TW) and N22 at the fixed [Fe/H] values are represented by solid and dashed lines, respectively. The slopes of the lines indicate the dependence on period terms, while the vertical separation between two lines from the different [Fe/H] level reflects the metallicity term's influence.}
\label{fig:pwz_ref}
\end{figure*}

Thus, we re-calibrate the PWZ relation for RRc stars with metallicities lower than [Fe/H] = $-1$, and list the results in Table\,\ref{tbl:fitcoff_PMZPWZ}. We then compare our fitting results with the recent study by \hyperlink{ngeow2022zwicky}{N22}, as illustrated in Figure\,\ref{fig:pwz_ref}. The fixed [Fe/H] values are set to $-$2 and $-$0.5 for RRab stars, and $-$2 and $-$1 for RRc stars. The clear agreement between the two studies in terms of both period and metallicity is evident, especially for the PW$_{gr}$Z relation of RRc stars after re-calibration. Moreover, our analysis indicates an even weaker dependence on metallicity, approaching zero in $W_{gr}$ for RRab stars, despite a higher scatter compared to Wesenheit magnitudes that include the $i$-band. This result is consistent with the theoretical work by \citet[][hereafter \hyperlink{muraveva2015new}{M15}]{muraveva2015new}, who also noted a weak metal dependency in the PWZ relations for $W_{BV}$ and $W_{BR}$. The consistency between our results and \hyperlink{muraveva2015new}{M15}'s work may be due to the significant overlap between the ZTF $gr$-bands transmission curves and those of the $BVR$ filters. Additionally, the trend in the coefficients for the period and metallicity terms across different Wesenheit functions in our PWZ relations also aligns with \hyperlink{muraveva2015new}{M15}'s theoretical predictions.

Utilizing this newly established PWZ relation, we estimate distances for our final sample. Due to the construction of the Wesenheit magnitude, which requires data from at least two bands, we can only derive distances for a subset of the ZTF\_RRL\_ALL sample. The distance uncertainties stem from two primary sources: (1) random errors estimated through 1000 MC simulations, which account for uncertainties in fit coefficients, photometric-metallicity, and apparent Wesenheit magnitude errors across the two bands used, and (2) methodological errors, evident from the scatter among various PWZ relations.

One sample can yield multiple distance estimates from different band relations. Therefore, Figures \ref{fig:pwz_dist_comself_ab} and \ref{fig:pwz_dist_comself_c} depict the internal comparisons across different bands for RRab and RRc stars, respectively. These comparisons reveal that both RRab and RRc stars exhibit acceptable agreements in terms of offsets and dispersions. Consequently, we provided reference distances using a weighted average method, similar to that employed in the PMZ relations. In total, we estimate distances for 64,156 RRLs in the ZTF\_RRL\_ALLsample: 46,185 RRab and 17,971 RRc stars. Only sample stars with photometric data from a single band have not had distances determined.

\section{Validation of Metallicity and Distance Estimates}\label{sec:vali}
In this section, the photometric-metallicity and distance estimates based on our newly constructed relations are compared to other estimates from the literature, and tests of the accuracy of these relations using GCs member stars are carried out.

\subsection{Validation with Recent Research}

Recently, \hyperlink{li2023photometric}{Li23} presented photometric-metallicity and distance estimates for over 130,000 RRLs using newly calibrated photometric-metallicity and M$_{G}-$[Fe/H] relations, based on Gaia DR3 data. For metallicity estimates, they compared their results with those from other photometric studies, such as \citet{dekany2022photometric}, showing an acceptable agreement when considering the differences in metallicity scales. Their results also align well with the high-resolution spectroscopic sample. Therefore, we compared our metallicity estimates with theirs, as shown in Appendix Figures \ref{metal_cpr_li_ab} (RRab) and \ref{metal_cpr_li_c} (RRc), using 2-D density maps similar to those in our previous internal comparison figures. Here, we also present results based on single-band relations, in addition to weighted average values, for validation across different bands. Overall, our work exhibits good agreement with \hyperlink{li2023photometric}{Li23}, with small offsets and acceptable scatter of 0.24 and 0.17\,dex, the maximum values in four comparisons for RRab and RRc stars, respectively. Specifically, a 0.9\,dex offset is found for the RRc stars, which modifies the systematic bias in \hyperlink{lli2023photometric}{Li23} when compared with GC members, as discussed in the next section. The slight tilts observed for RRab and RRc stars, notably on the metal-rich side, is mainly attributed to differences in boundary value selection during the calibration of the relationship, due to varying sample sizes.

\begin{figure}[!htbp]
\centering
\includegraphics[width=0.98\linewidth]{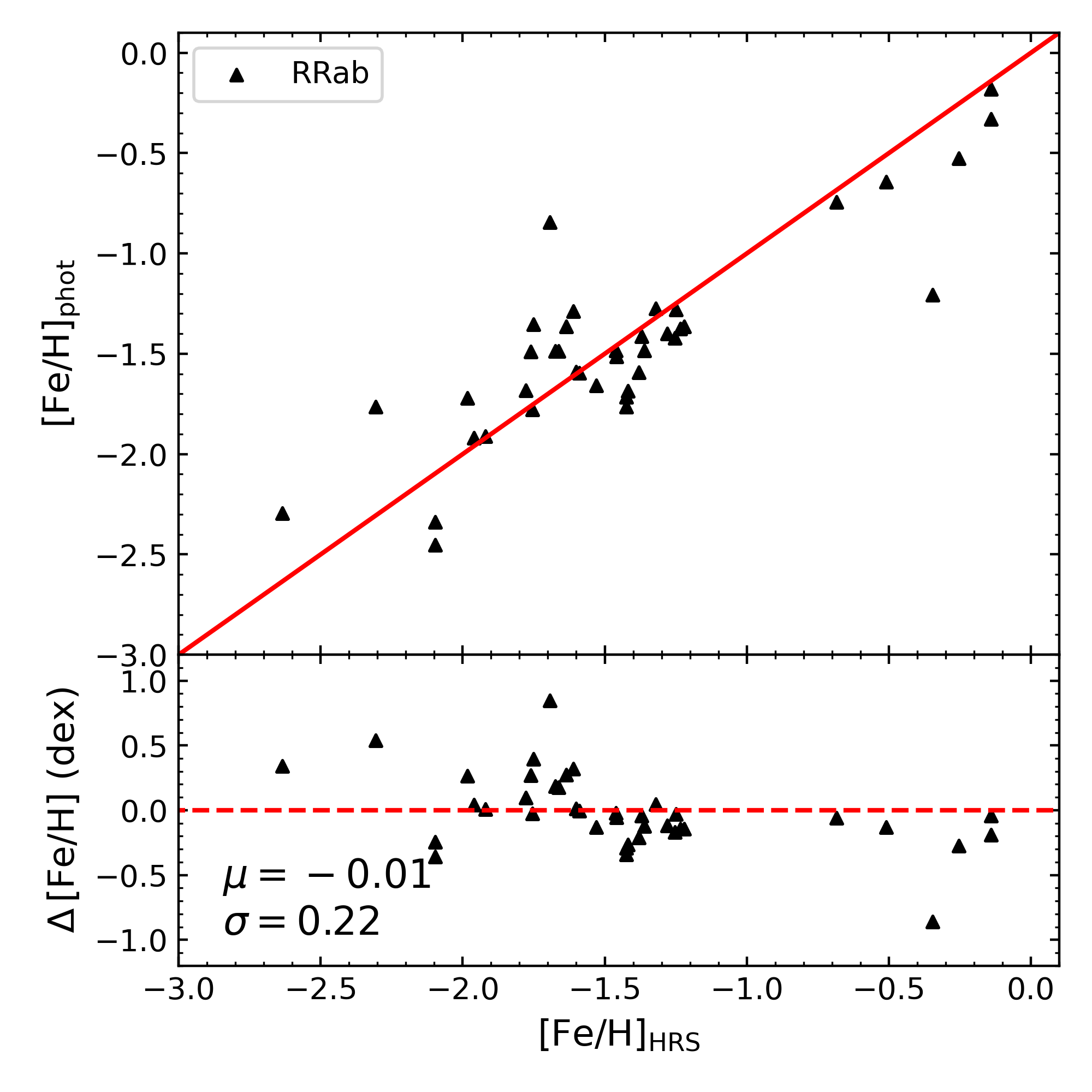}
\caption{Upper section of panel: Comparison of photometric-metallicity estimates from this work and metallicity from the high-resolution spectroscopic sample. The red-solid line is the one-to-one line. Lower section of panel: 
$\Delta$[Fe/H] differences (our work minus high-resolution spectroscopic results), with the mean and standard deviation displayed in the lower left. The red-dashed line is the zero residual level.}
\label{fig:metal_hrs}
\end{figure}

We now assess the accuracy of our photometric-metallicity estimates using the high-resolution spectroscopic sample.  
\citet{dekany2021metallicity} integrated 183 RRab and 49 RRc stars with calibrated, same-scale metallicity measurements from high-resolution spectroscopy, covering the range from [Fe/H ] = $-$3.1 to +0.2, providing the most comprehensive high-resolution spectroscopic sample to date. Cross-matching with this sample, we identified 39 common RRab stars, but fewer than 10 common RRc stars with large errors due to the limited total number of RRc sample stars. To ensure the credibility of our comparison, we conducted the comparison only for RRab stars, as shown in Figure\,\ref{fig:metal_hrs}. For type RRab stars, the metallicities from this work are in excellent agreement with those from high-resolution spectroscopy, with a tiny offset of $-$0.01\,dex and a dispersion of 0.22\,dex.

In addition to validating metallicity, we also present comparisons of distance estimates derived from the PMZ and PWZ relations with those of \hyperlink{li2023photometric}{Li23}. Appendix Figures \ref{dist_pmz_li_ab} and \ref{dist_pmz_li_c} compare our PMZ-based results against \hyperlink{li2023photometric}{Li23}'s findings for RRab and RRc stars, respectively.  Figure\,\ref{dist_pmz_li_ab}, which excludes sample stars using A$_G$ from \hyperlink{clementini2023gaia}{C23}, shows that our results align closely with \hyperlink{li2023photometric}{Li23}, with a minor offset and an acceptable level of scatter, as seen in the residual plots. The concordance is even more pronounced for RRc stars.

Appendix Figures \ref{dist_pwz_li_ab} and \ref{dist_pwz_li_c} compare our PWZ-based results against \hyperlink{li2023photometric}{Li23}'s findings for RRab and RRc stars, respectively.
Significant scatter in the W$_{gr}$-bands are noted, for both RRab and RRc stars, especially for sample stars with distances between 8 and 11\,kpc. For further analysis, we divided these sample into two categories: `normal' and `strange', based on their scatter levels relative to more distant sample stars. We then examined their positions in the Galactic coordinate system. Upon analysis, we found that the `normal' sample stars had a similar sky distribution to the total sample, while the `strange' sample stars were predominantly concentrated near the Galactic center, specifically within the regions where $|b| \leq 15^{\circ}$ and $-10^{\circ} \leq |l| \leq 30^{\circ}$.

The consistent performance across all bands for the distance estimates, coupled with the high-extinction environment of these sample stars, suggests that the dispersion is largely attributable to differing extinction sources. In the high-extinction region, the distances ($d_G$) for RRab stars from \hyperlink{li2023photometric}{Li23} utilized the A$_{G}$ values derived from the empirical relation of \hyperlink{clementini2023gaia}{C23}, which considers the period, amplitude of the $G$-band, and two observational apparent magnitudes. This method showed a two-sided scatter when compared to our $d_{gr}$ results. For RRc stars, the $d_G$ using the A$_{G}$ derived from the \hyperlink{schlegel1998maps}{SFD98} map exhibited a one-sided scatter compared to our $d_{gr}$ estimates. However, in these comparison, both the empirical relation and the \hyperlink{schlegel1998maps}{SFD98} map are subject to uncertainties due to the high-extinction environment. Our reliance on Wesenheit magnitudes is constrained by the assumed universal reddening law and the precision of the mean magnitudes, potentially leading to biases due to high-extinction environments. Consequently, it is difficult to judge the accuracy of these sample stars' distances or to assess which method is more reliable. However, aside from a few ($<$ 5\%) outliers, the overall comparison of the remaining sample stars demonstrates good consistency.

\begin{figure}[!htbp]
\centering
\includegraphics[width=0.98\linewidth]{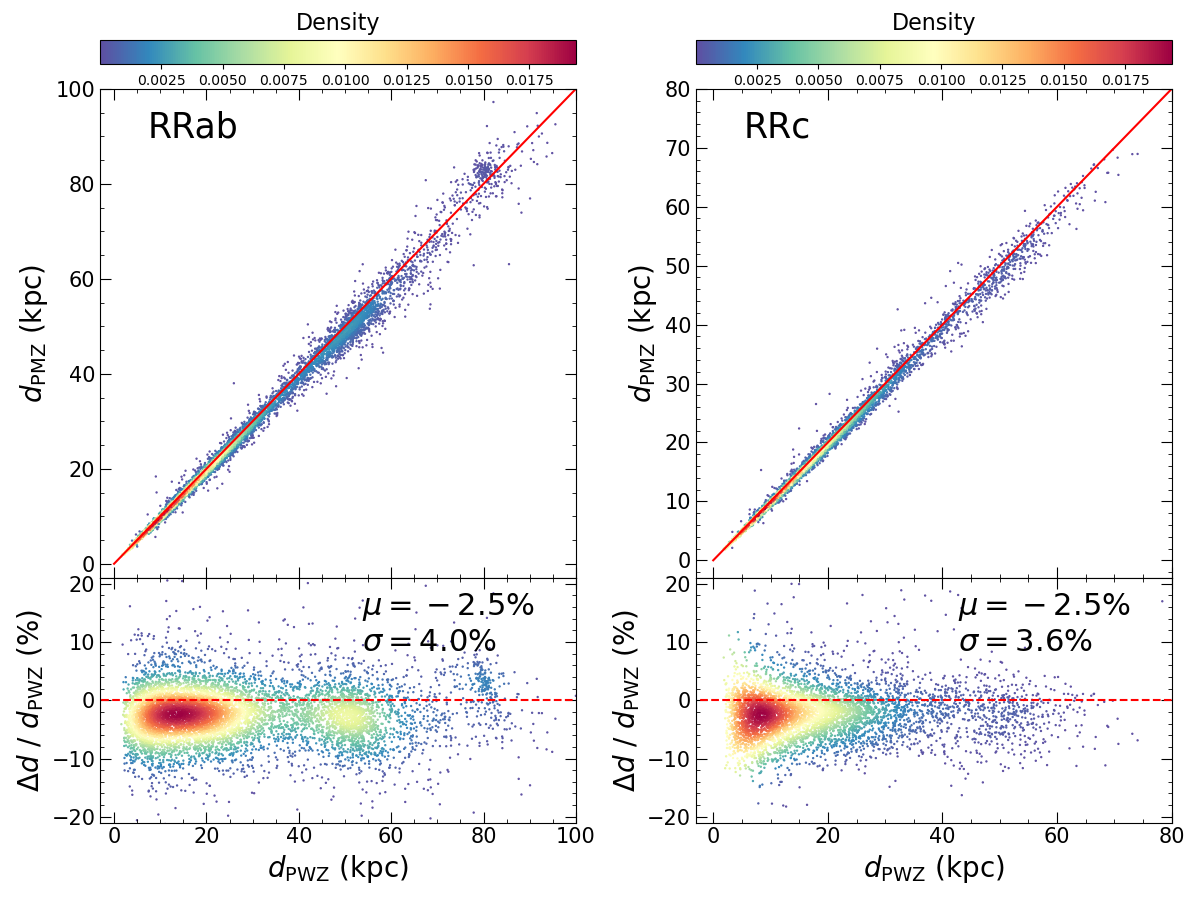}
\caption{Comparison of distance estimates calculated from our PMZ and PWZ relations for RRab stars 
(left panel) and RRc stars (right panel). The solid-red line is the one-to-one line. The lower section of each panel displays the relative difference between 
$d_\text{PMZ}$ and $d_\text{PWZ}$, along with the mean and standard deviation of the relative difference indicated at the upper right of each panel. The dashed-red line is the zero residual level.}
\label{fig:pwz_pmz_comp}
\end{figure}

Considering the two concurrent distance estimates for each sample based on the PMZ and PWZ relations, we compare them using common sample stars, as illustrated in Figure\,\ref{fig:pwz_pmz_comp}. Note that potential low-quality sample stars and RRab stars with $|b| \leq 25^{\circ}$ are excluded from the comparison,  to more accurately reflect the differences between the two calibration methods. A mild negative offset is clearly observed for ${d_\text{PMZ}}$ when compared with ${d_\text{PWZ}}$, indicating generally lower distance estimates. Meanwhile, a mild scatter in the $\Delta d/d$ also suggests that the differences may primarily be related to extinction values. Therefore, we provide both distance values, and give priority to one after validations with globular clusters, as discussed below.

\subsection{Validation with GCs}\label{sec:vali_gcs}
GCs member stars typically share similar metallicities due to a common formation history. Thus, we utilize RRLs from GCs to verify the accuracy of our newly calibrated photometric-metallicity and distance relations. Our test samples are sourced from two parts: the primary portion consists of member stars from the ZTF\_RRL\_FIT sample, while the supplementary portion comes from the catalog of \hyperlink{ngeow2022zwicky}{N22}'s work. 

For the primary portion, we apply selection criteria similar to those of \hyperlink{li2023photometric}{Li23} to 157 GCs from \citet[][hereafter \hyperlink{harris2010new}{H10}]{harris2010new}. We use the proper motions and their uncertainties for each GC from \citet[][hereafter \hyperlink{vasiliev2021gaia}{BV21}]{vasiliev2021gaia}. The criteria include:

\begin{itemize}[leftmargin=*]

\item The angular distance to the GC center must be less than 15 half-light radii $r_{\rm h}$

\item The proper motions must satisfy $|\mu_{\alpha} - \mu_{\alpha, {\rm GC}}| \leq 8\, \sigma_{\mu_{\alpha, {\rm GC}}}$ and\,$|\mu_{\delta} - \mu_{\delta, {\rm GC}}| \leq 8\, \sigma_{\mu_{\delta, {\rm GC}}}$

\end{itemize}

For the supplementary portion, we selected sample stars not flagged with `ACR' from Table\,3 of \hyperlink{ngeow2022zwicky}{N22}, and re-evaluated those not previously labeled using our previously discussed method. In total, we obtained 417 sample stars, comprising 272 RRab and 145 RRc stars. To ensure confident validation of GCs, a quota of at least five RRab or three RRc stars is necessary for selecting members of each cluster. Then, the mean metallicity and distance, along with their uncertainties, are calculated for each GC.

\begin{figure}[!htbp]
\centering
\includegraphics[width=1\linewidth]{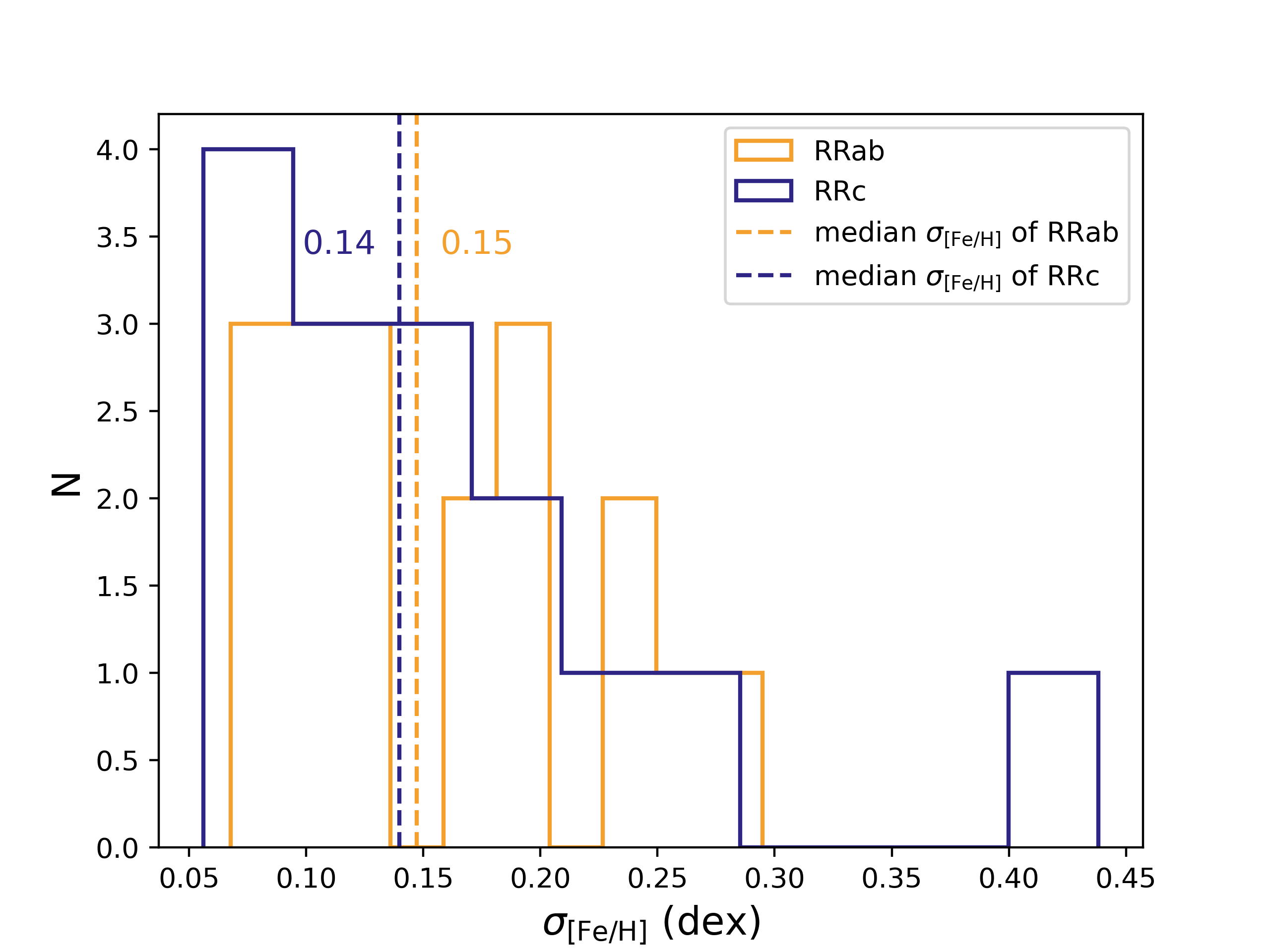}
\caption{The uncertainty distributions of our metallicity estimates for each globular cluster are shown for RRab (blue) and RRc (yellow) member stars, respectively. The typical accuracies in median values are indicated with dashed-blue and dashed-yellow lines for RRab and RRc stars,
respectively.}
\label{fig:clu_feh_sig}
\end{figure}

\begin{figure}[!htbp]
\centering
\includegraphics[width=0.98\linewidth]{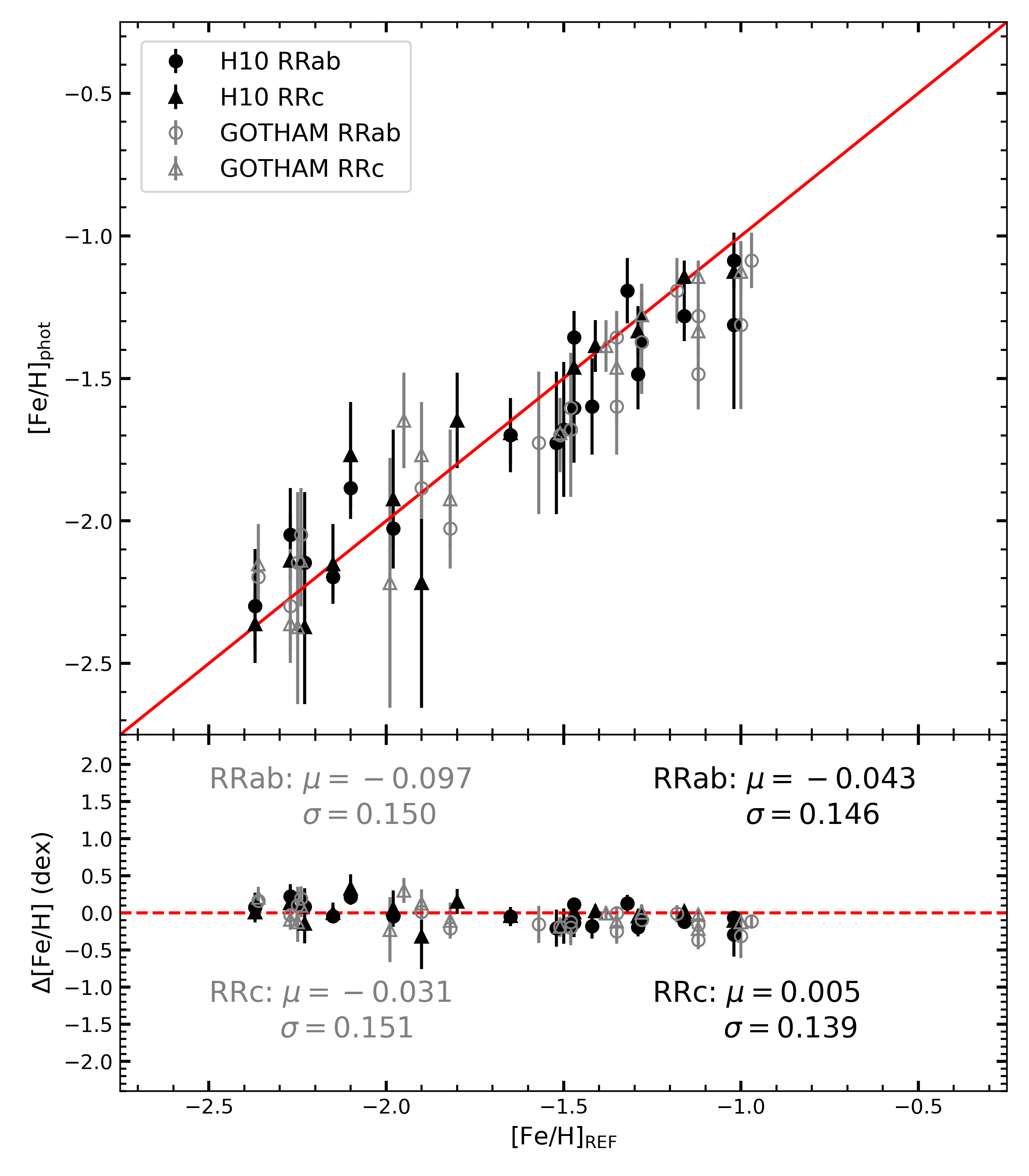}
\caption{Our estimated metallicity (y-axis) compared with two reference works (x-axis), with black dots representing H10 and gray dots representing the GOTHAM survey. Circles denote RRab stars, and triangles denote RRc stars. The solid-red line is the one-to-one line. The mean and scatter of the residuals are shown in the lower panel.  The dashed-red line is the zero residual level.}
\label{fig:clu_feh_comp}
\end{figure}

As shown in Figure\,\ref{fig:clu_feh_sig}, the typical uncertainties in metallicity ([Fe/H]\_wei\_TW) in this study are 0.15\,dex for RRab stars in 18 GCs and 0.14\,dex for RRc stars in 15 GCs. Table\,\ref{tbl:GCs_metaldist} lists the mean and uncertainty of the metallicity estimates, the member numbers used for calibration for each GC, as well as referenced metallicities from \hyperlink{harris2010new}{H10} and the GlObular clusTer Homogeneous Abundances Measurements
(GOTHAM) survey (\citealt{dias2015fors2,dias2016fors2,dias2016globular,vasquez2018homogeneous}). Our photometric-metallicity estimates, compared with referenced works, are shown in Figure\,\ref{fig:clu_feh_comp}. Our results closely match those of \hyperlink{harris2010new}{H10}, with a small offset of $-$0.04\,dex and a scatter of 0.15\,dex for RRab stars, and a negligible offset of 0.005\,dex and a scatter of 0.14\,dex for RRc stars. In contrast, \hyperlink{li2023photometric}{Li23} reports larger offsets of $-$0.09 and $-$0.12\,dex, and larger scatters of 0.15 and 0.16\,dex for RRab and RRc stars, respectively. Our performance is even superior to \hyperlink{li2023photometric}{Li23} when compared with \hyperlink{harris2010new}{H10}.

\begin{table*}
\centering
\caption{Comparison of Photometric-metallicities and Distance Estimates with the Reference Values for GCs\label{tbl:GCs_metaldist}}
\begin{threeparttable}
\begin{tabular}{ccccccccccc}

\hline\hline
Name&[Fe/H]$_{\rm H10}$&[Fe/H]$_{\rm GOTHAM}$&[Fe/H]$_{\rm TW}$&N$_{\rm [Fe/H]}$&$d_{\rm H10}$&$d_{\rm BV21}$&$d_{\rm PMZ}$&N$_{\rm PMZ}$&$d_{\rm PWZ}$&N$_{\rm PWZ}$\\
& & & & &(kpc)&(kpc)&(kpc)&&(kpc)&\\

\hline
 \noalign{\smallskip}
\multicolumn{11}{c}{RRab} \\ \hline
 \noalign{\smallskip}
 
 NGC 4590&$-2.23$&$-2.25\pm0.02$&$-2.15\pm0.25$&10&$10.3$&$10.4\pm0.10$&$9.89\pm0.49$&10&$9.99\pm0.49$&10\\
 NGC 5024&$-2.10$&$-1.90\pm0.05$&$-1.89\pm0.11$&17&$17.9$&$18.50\pm0.18$&$17.58\pm0.21$&17&$18.81\pm0.27$&16\\
 NGC 5053&$-2.27$&$-2.24\pm0.16$&$-2.05\pm0.16$&5&$17.4$&$17.54\pm0.23$&$16.49\pm0.32$&5&$17.26\pm0.27$&5\\
 NGC 5272&$-1.50$&$-1.48\pm0.05$&$-1.68\pm0.24$&21&$10.2$&$10.18\pm0.08$&$10.05\pm0.15$&21&$10.39\pm0.16$&16\\
 NGC 5466&$-1.98$&$-1.82\pm0.08$&$-2.03\pm0.07$&11&$16.0$&$16.12\pm0.16$&$15.45\pm0.20$&11&$15.98\pm0.17$&11\\
 NGC 5904&$-1.29$&$-1.12\pm0.01$&$-1.49\pm0.12$&27&$7.5$&$7.48\pm0.06$&$7.23\pm0.21$&27&$7.65\pm0.21$&27\\
 NGC 6121&$-1.16$&$-1.12\pm0.02$&$-1.28\pm0.09$&5&$2.2$&$1.85\pm0.02$&$2.00\pm0.20$&5& --&--\\
 NGC 6171&$-1.02$&$-1.00\pm0.02$&$-1.31\pm0.29$&8&$6.4$&$5.63\pm0.08$&$5.92\pm0.27$&8&$5.39\pm0.08$&5\\
 NGC 6229&$-1.47$&$-1.35\pm0.16$&$-1.36\pm0.09$&11&$30.5$&$30.11\pm0.47$&$28.93\pm0.67$&11&$29.78\pm0.69$&11\\
 NGC 6402&$-1.28$&$-1.28\pm0.05$&$-1.37\pm0.18$&13&$9.3$&$9.14\pm0.25$&$8.19\pm0.33$&13&$8.57\pm0.37$&12\\
 NGC 6426&$-2.15$&$-2.36\pm0.03$&$-2.20\pm0.07$&7&$20.6$&$20.71\pm0.35$&$20.01\pm0.53$&7&$19.60\pm0.68$&7\\
 NGC 6626&$-1.32$&$-1.18\pm0.05$&$-1.19\pm0.11$&5&$5.5$&$5.37\pm0.10$&$5.30\pm0.16$&5&$5.65\pm0.54$&5\\
 NGC 6712&$-1.02$&$-0.97\pm0.05$&$-1.09\pm0.10$&6&$6.9$&$7.38\pm0.24$&$6.81\pm0.29$&6&$7.75\pm1.14$&6\\
 NGC 6934&$-1.47$&$-1.48\pm0.11$&$-1.60\pm0.19$&29&$15.6$&$15.72\pm0.17$&$14.88\pm0.55$&29&$15.47\pm0.79$&24\\
 NGC 6981&$-1.42$&$-1.35\pm0.08$&$-1.60\pm0.17$&20&$17.0$&$16.66\pm0.18$&$16.26\pm0.41$&20&$17.21\pm0.46$&19\\
 NGC 7006&$-1.52$&$-1.57\pm0.05$&$-1.73\pm0.25$&13&$41.2$&$39.32\pm0.56$&$40.79\pm1.61$&13&$40.87\pm1.61$&12\\
 NGC 7078&$-2.37$&$-2.27\pm0.01$&$-2.30\pm0.20$&10&$10.4$&$10.71\pm0.10$&$10.32\pm0.41$&10&$10.20\pm0.38$&10\\
 NGC 7089&$-1.65$&$-1.51\pm0.02$&$-1.70\pm0.13$&5&$11.5$&$11.69\pm0.11$&$10.93\pm0.30$&5& --& --\\

\hline
 \noalign{\smallskip}

\multicolumn{11}{c}{RRc} \\ \hline
 NGC 4147&$-1.80$&$-1.95\pm0.09$&$-1.65\pm0.17$&4&$19.3$&$18.54\pm0.21$&$17.82\pm0.77$&4&$18.36\pm1.10$&4\\
 NGC 4590&$-2.23$&$-2.25\pm0.02$&$-2.37\pm0.27$&17&$10.3$&$10.4\pm0.10$&$10.19\pm0.40$&17&$10.00\pm0.33$&16\\
 NGC 5024&$-2.10$&$-1.90\pm0.05$&$-1.77\pm0.19$&10&$17.9$&$18.50\pm0.18$&$17.68\pm0.43$&10&$19.01\pm0.45$&9\\
 NGC 5053&$-2.27$&$-2.24\pm0.16$&$-2.14\pm0.16$&4&$17.4$&$17.54\pm0.23$&$16.72\pm0.51$&4&$17.67\pm0.15$&4\\
 NGC 5466&$-1.98$&$-1.82\pm0.08$&$-1.92\pm0.24$&5&$16.0$&$16.12\pm0.16$&$15.44\pm0.65$&5&$16.23\pm0.46$&5\\
 NGC 5897&$-1.90$&$-1.99\pm0.03$&$-2.33\pm0.44$&4&$12.5$&$12.55\pm0.24$&$12.30\pm0.88$&4&--&--\\
 NGC 5904&$-1.29$&$-1.12\pm0.01$&$-1.33\pm0.09$&12&$7.5$&$7.48\pm0.06$&$7.20\pm0.81$&12&$7.95\pm0.76$&9\\
 NGC 6121&$-1.16$&$-1.12\pm0.02$&$-1.14\pm0.06$&4&$2.2$&$1.85\pm0.02$&$2.07\pm0.09$&4&$1.90\pm0.03$&4\\
 NGC 6171&$-1.02$&$-1.00\pm0.02$&$-1.13\pm0.06$&6&$6.4$&$5.63\pm0.08$&$5.68\pm0.16$&6&$5.31\pm0.05$&6\\
 NGC 6229&$-1.47$&$-1.35\pm0.16$&$-1.46\pm0.20$&5&$30.5$&$30.11\pm0.47$&$29.31\pm0.27$&5&$30.39\pm0.48$&5\\
 NGC 6402&$-1.28$&$-1.28\pm0.05$&$-1.28\pm0.11$&9&$9.3$&$9.14\pm0.25$&$8.28\pm0.44$&9&$8.66\pm0.24$&8\\
 NGC 6426&$-2.15$&$-2.36\pm0.03$&$-2.15\pm0.14$&4&$20.6$&$20.71\pm0.35$&$19.65\pm0.46$&4&$19.56\pm0.64$&4\\
 NGC 7078&$-2.37$&$-2.27\pm0.01$&$-2.36\pm0.10$&14&$10.4$&$10.71\pm0.10$&$10.42\pm0.15$&14&$10.67\pm0.12$&14\\
 NGC 7089&$-1.65$&$-1.51\pm0.02$&$-1.69\pm0.11$&5&$11.5$&$11.69\pm0.11$&$10.95\pm0.21$&5&$12.53\pm0.54$&4\\
 Pal 5&$-1.41$&$-1.38\pm0.16$&$-1.39\pm0.09$&8&$23.2$&$21.94\pm0.51$&$21.02\pm0.42$&8&$21.22\pm0.42$&8\\

\hline
\end{tabular}
\tablecomments{Column(1): Cluster identification number; Columns(2) and (3): Metallicities from \hyperlink{harris2010new}{H10} and the GOTHAM suvery; Columns(4) and (5): Photometric metallicities of GCs with uncertainties estimated by this work and the member counts used for the calculations; Columns (6) and (7): Distances from \hyperlink{harris2010new}{H10} and \hyperlink{vasiliev2021gaia}{BV21}; Columns (8) and (9): Weighted average distance based on our newly calibrated PMZ relation and the member counts used for the calculations; Columns (10) and (11): Weighted average distance based on our newly calibrated PWZ relation and the member counts used for the calculations.}
\end{threeparttable}
\end{table*}

\begin{figure*}[!htbp]
\centering
\includegraphics[width=0.98\linewidth]{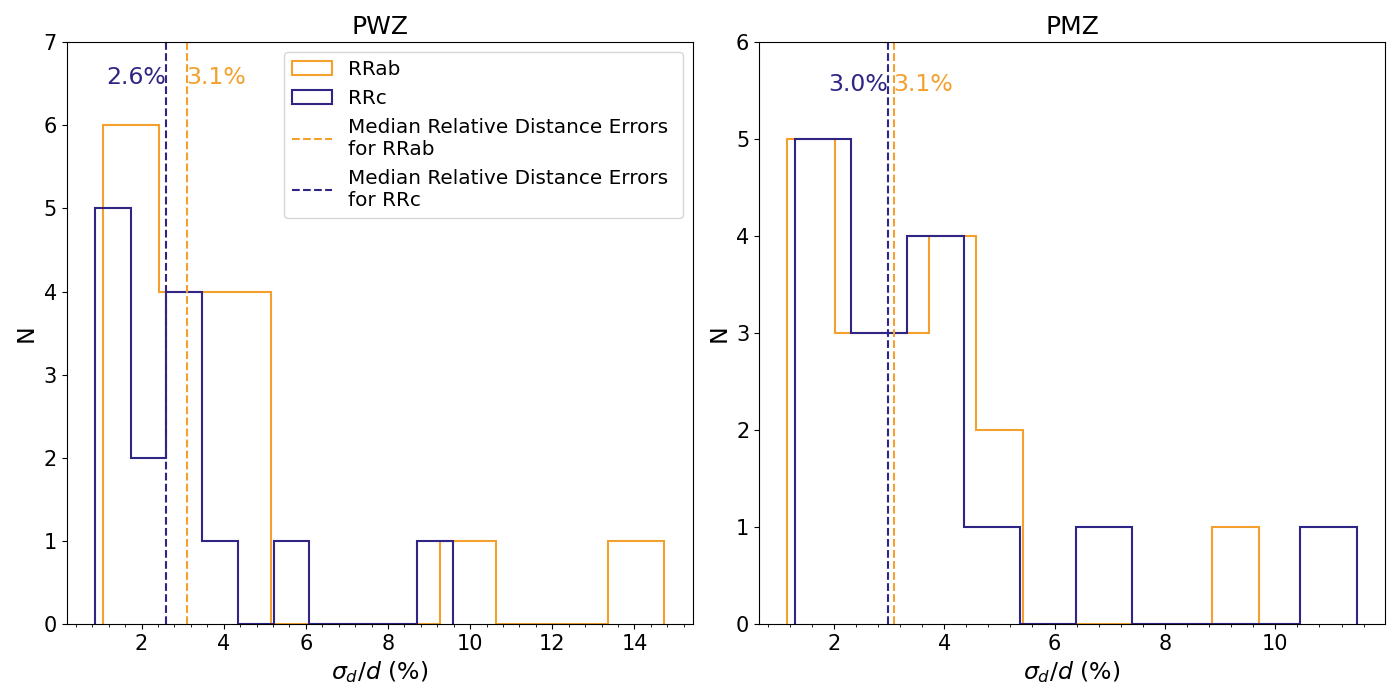}
\caption{Distribution of relative distance errors for the GCs.  Left panel: Using weighted average distances calculated with the newly calibrated PWZ relations.  Right panel: Using the PMZ relations. The median relative errors of distance are indicated with dashed-blue and dashed-yellow lines for RRab and RRc stars, respectively.}
\label{fig:clu_pmzpwz_sig}
\end{figure*}

For the distance validation, we assessed the accuracy of the weighted average distances derived from the PMZ and PWZ relations, as shown in Figure\,\ref{fig:clu_pmzpwz_sig}. Using the PMZ relation, we validated the distances for 18 GCs for RRab and 15 GCs for RRc stars. The relative distance errors for RRab and RRc stars were mostly within 6\% and 8\%, with median errors of 3.1\% and 3.0\%. With the PWZ relation, we validated the distances for 16 GCs for RRab and 14 GCs for RRc stars. The relative distance errors for RRab and RRc stars were mostly within 10\% and 6\%, with median errors of 3.1\% and 2.6\%, respectively. Additional results for single-band or band-pair relations are shown in Appendix Figure\,\ref{fig:clu_dist_6sig}. The calculated distance are listed in Table\,\ref{tbl:GCs_metaldist}, and summarized in Table\,\ref{tbl:clu_acc_dist}.

\begin{table}
\centering
\caption{Summary of Accuracies for the GC Distance Estimates\label{tbl:clu_acc_dist}}
\begin{threeparttable}
\setlength{\tabcolsep}{8mm}{
\begin{tabular}{lcc}
\hline
\begin{minipage}{5mm}\vspace{4.5mm} \vspace{1mm} \end{minipage} 
Distance & $\sigma_{d}/d$ (\%)\tnote{a} & N$_\text{GCs}$\tnote{b} \\ 
\hline
$d_{g}$  & 3.2 /3.9 & 17 /15\\
$d_{r}$  & 2.6 /2.6 & 16 /15\\
$d_{i}$  & 1.5 /2.1 & 9 /7\\
$d_{gr}$ & 3.1 /2.4 & 16 /14\\
$d_{gi}$ & 1.7 /2.1 & 9 /7\\
$d_{ri}$ & 1.9 /2.2 & 9 /7\\
$d_{\rm PMZ}\tnote{c}$ & 3.1 /3.0 & 18/15\\
$d_{\rm PWZ}\tnote{d}$ & 3.1 /2.6 & 16/14\\
\hline
\end{tabular}}
\begin{tablenotes}
\item[a] Relative-distance errors are shown, with left values for RRab stars and right values for RRc stars.
\item[b] Number of GCs used for distance-accuracy validation, with left values for RRab stars and right values for RRc stars.
\item[c] Weighted average distances based on PMZ relations across the $gri$-bands.
\item[d] Weighted average distances based on PWZ relations across $gr$-, $gi$-, and $ri$-band pairs.
\end{tablenotes}
\end{threeparttable}
\end{table}

We also compared our distances with those of \hyperlink{vasiliev2021gaia}{BV21} and \hyperlink{harris2010new}{H10}, as shown in Appendix Figure\,\ref{clu_dist_pmzpwz_comp}. The distances used from \hyperlink{vasiliev2021gaia}{BV21} represent the mean of uniform scale distances derived from multiple independent measurements. In contrast, \hyperlink{harris2010new}{H10}'s distances are primarily based on the mean $V$ magnitude of the horizontal branch of GCs provided by the references. Based on the mean offsets and scatters of relative distance differences, our derived distances for GCs exhibit better agreement with \hyperlink{vasiliev2021gaia}{BV21} than with \hyperlink{harris2010new}{H10}. This is particularly true for the PWZ relations, which demonstrate negligible mean offsets of $-$0.4\% for RRab and $-$0.3\% for RRc stars, along with small scatters of 3.7\% and 4.0\%, respectively. Similar scatters are observed in the PMZ relations, but mild overall offsets are evident, with $-$2.7\% for RRab and $-$3.9\% for RRc stars. Considering the small $E(B-V)$ values of GCs, this suggests that the PMZ relation may have systematic offsets during its calibration process. Therefore, the recommended choice for distance estimates is to use values calculated from PWZ relations for sample stars that have two distinct distance estimates. Similarly, additional comparisons for single-band or band-pair relations are shown in Appendix Figure\,\ref{fig:clu_dist_6comp}.

\section{Final Catalog}\label{sec:cat}

\begin{figure}[!htbp]
\centering
\includegraphics[width=1\linewidth]{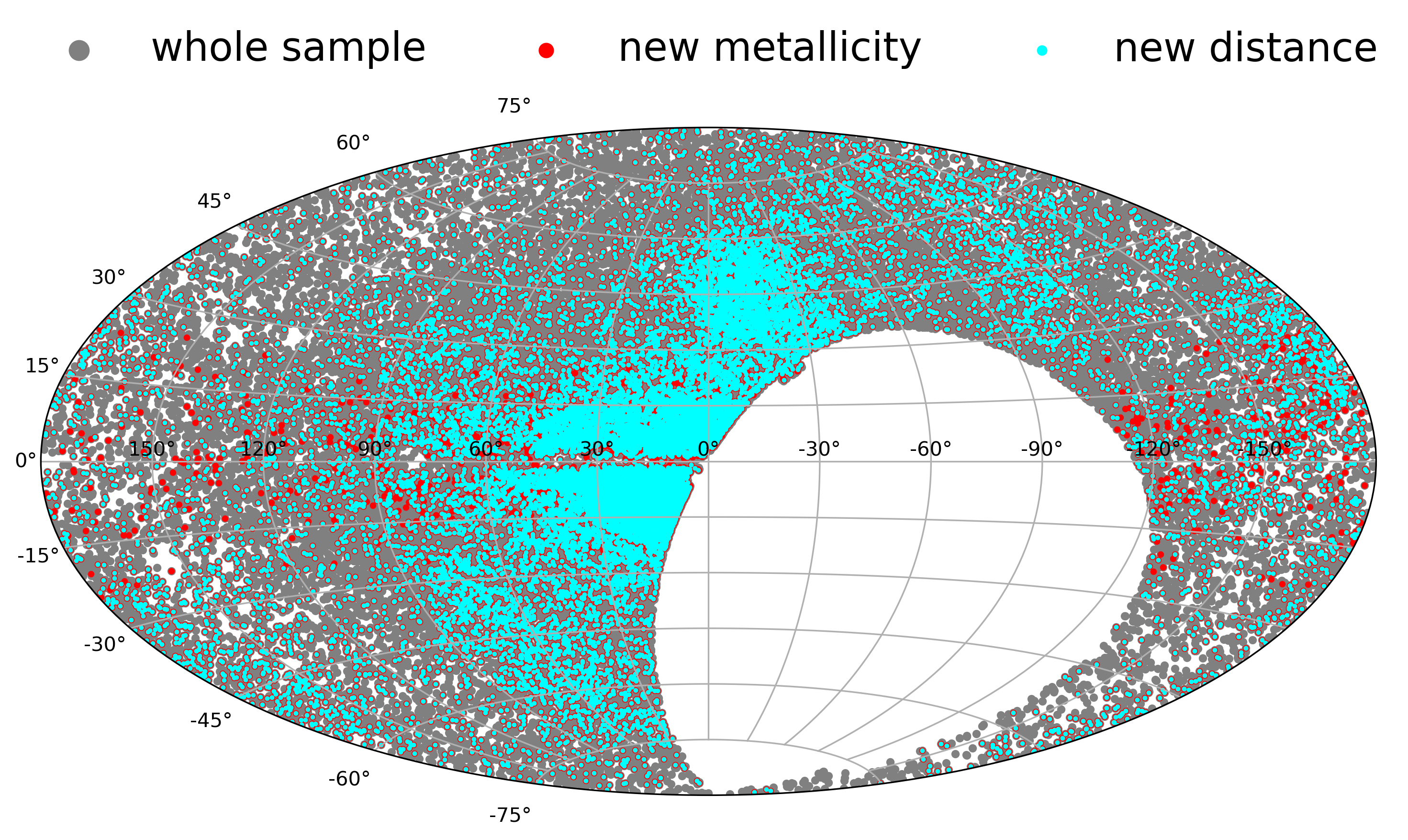}
\caption{Sky distribution in Galactic coordinates of our final sample are shown in gray, with the sample stars with newly obtained photometric-metallicity estimates colored red, and those with newly obtained distance estimates colored cyan.}
\label{fig:sky_distri}
\end{figure}

Our final catalog includes the ZTF\_RRL\_ALL sample, comprising 73,795 sample stars (52,571 RRab and 21,224 RRc stars) with photometric-metallicity estimates from the newly calibrated $P-\phi_{31}-R_{21}-\text{[Fe/H]}$ and $P-\phi_{31}-A_{2}-A_{1}-\text{[Fe/H]}$ relations. Over 95\% of the sample stars (70,560 in total; 52,050 RRab stars and 18,510 RRc stars) have accurate distance measurements from our PMZ/PWZ relations. By cross-matching with \hyperlink{li2023photometric}{Li23}'s work, we derived 25,439 sample stars (34\% of the total sample) that filled in the metallicity parameters and 22,937 sample stars (33\% of the total sample) that filled in the distance parameters. The sky distribution of the entire sample, as well as those with initially derived metallicity and distance, is shown in Figure\,\ref{fig:sky_distri}. 

\begin{figure}[!htbp]
\centering
\includegraphics[width=0.98\linewidth]{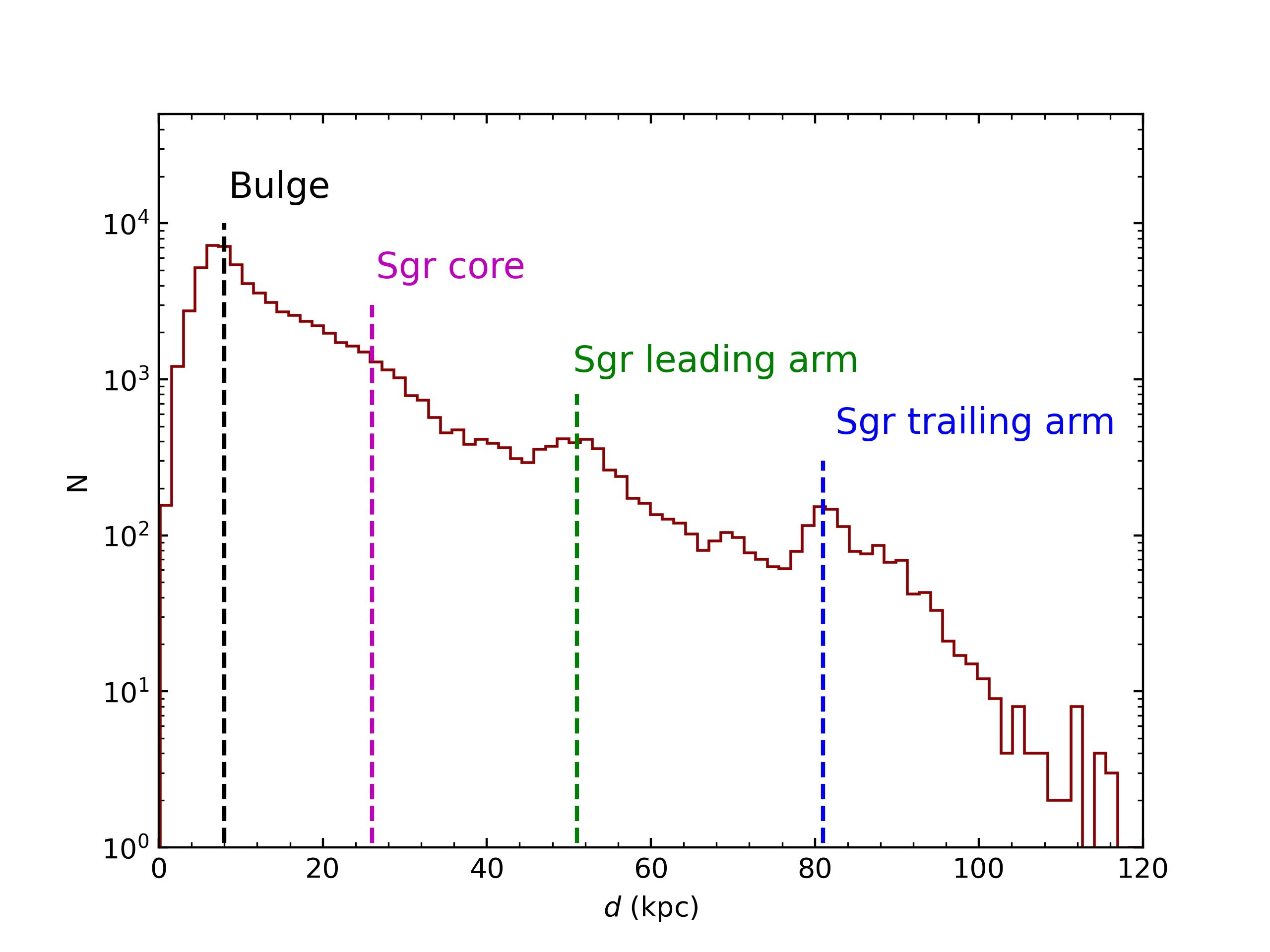}
\caption{The heliocentric distance distribution for our final RRL sample. The prominent structures are marked by dashed lines in different colors. The y-axis represents the frequency on a logarithmic scale.}
\label{fig:dist_distri}
\end{figure}

The distance distribution of our final sample extends to over 100\,kpc, with four prominent peaks visible as a function of heliocentric distances, as shown in Figure\,\ref{fig:dist_distri}. The peak at 8\,kpc corresponds to the Galactic bulge, the peaks at 26, 51, and 81\,kpc correspond to distinct segments of the Sagittarius core, leading arm and trailing arm (\citealt{purcell2011sagittarius,fardal2019,ramos2022,Sun2024}). The relatively large increment in sample size and acceptable accuracy of physical parameters will facilitate a more in-depth understanding of our Galaxy's structure, as well as its chemical and kinematic properties. Our final catalog, presented in Table\,\ref{tbl:fincat}, outlines the column names that are part of the online sample catalog. This catalog is also accessible on Zenodo\footnote{\href{https://doi.org/10.5281/zenodo.14561442}{10.5281/zenodo.14561442}}.

\section{Summary}\label{sec:summ}

In this work, we have selected 73,796 relatively high-quality ZTF RR Lyrae stars from multiple catalogs, including those from Gaia, ZTF, ASAS-SN, and PS1. Utilizing re-calculated periods and Fourier parameters from sample stars in the $gri$-bands, we established $P-\phi_{31}-R_{21}-\text{[Fe/H]}$ and $P-\phi_{31}-A_{2}-A_{1}-\text{[Fe/H]}$ relations in the ZTF photometric system for RRab and RRc stars, respectively, benefiting from the spectral sample of 2,875 RRab and 1,182 RRc stars provided by \hyperlink{Liu2020}{L20}. After validation across different bands, we found strong internal consistency, and derived photometric-metallicity estimates for 73,795 RRLs (34\% derived for the first time, compared with \hyperlink{li2023photometric}{Li23}), using a weighted average method. External comparisons with \hyperlink{li2023photometric}{Li23} and the high-resolution spectroscopic sample exhibit no significant offsets, with typical precisions of 0.15\,dex and 0.14\,dex for RRab and RRc stars, respectively, after validation using GCs members.

Using hundreds of local bright RRLs with metallicity estimates from this work and accurate distance estimates from Gaia parallaxes, we have re-calibrated the PMZ and PWZ relations in the $gri$-bands. Given the good agreement observed in internal comparisons, we subsequently applied the weighted average method to both relations. Subsequently, we derived distance estimates for 56,593 RRLs using the PMZ relation and 64,156 RRLs using the PWZ relation. The results of our calibration and external validation are summarized as follows:

  1. For the calibration of the PMZ relations in the $gri$-bands, the change in the fitted coefficients with wavelength follows a similar trend as in previous works. However, the slope of the period term in the $g$-band differs significantly from two reference works, while the $r$- and $i$-bands are more consistent. A possible explanation is the different prior settings for the period-term coefficient, which may be more sensitive during the fitting process.

  2. In the section on the calibration of the PWZ relations, we unexpectedly find a small W$_{gr}$ for metal-rich RRc sample stars. After excluding these stars, the fitting results exhibit good agreement with \hyperlink{ngeow2022zwicky}{N22}, and the trend in the coefficients for the period and metallicity terms across different Wesenheit functions aligns with the theoretical work of \hyperlink{muraveva2015new}{M15}.

  3. When compared with the \hyperlink{li2023photometric}{Li23} distance estimates, the PMZ relations exhibit minor offsets and small scatters for $\Delta d/d$ in both RRab and RRc stars. The PWZ relations show similar offsets and acceptable scatter when excluding the `strange' sample stars.

  4. In cluster validations, the typical relative errors of the weighted average distance from the PMZ relation are 3.1\% and 3.0\% for RRab and RRc stars, respectively, and 3.1\% and 2.6\% from the PWZ relation, respectively. Compared to the distances provided by \hyperlink{vasiliev2021gaia}{BV21}, the PWZ relation shows better agreement, and is recommended as the preferred choice. With the upcoming release of Gaia DR4, it is anticipated that calibrating a larger sample of nearby field stars will help to improve the systematic bias in the PMZ relation.

Ultimately, by integrating distance estimates from both PMZ and PWZ relations, we obtained estimates for 70,560 RRLs, including 22,937 previously unmeasured in Gaia DR3. As future time-domain surveys, such as LSST and SiTian, emerge, we anticipate the discovery of more RRLs and the derivation of precise physical parameters, significantly enhancing our understanding of the Galaxy and universe.

\section*{Acknowledgements} 

We appreciate the anonymous referee’s valuable comments, which helped us improve the paper. This work acknowledges support from the National Natural Science Foundation of China (NSFC) for grants Nos. 12090041, 12422303, and 12090040, the support by National Key R\&D Program of China (Grant No. 2023YFA1608303) and the China Manned Space Project with No. CMS-CSST-2021-A08. This work is supported by the Strategic Priority Research Program of the Chinese Academy of Sciences (grant Nos. 
XDB0550100, XDB0550000, and XDB0550103). G.C.L. acknowledges support from the Hubei Provincial Natural Science Foundation with grant no. 2023AFB577, and the National Science Foundation of China (NSFC) with grant no. U1731108. T.C.B. acknowledges partial support for this work from grant PHY 14-30152; Physics Frontier Center/JINA Center for the Evolution of the Elements (JINA-CEE), and OISE-1927130: The International Research Network for Nuclear Astrophysics (IReNA), awarded by the US National Science Foundation.

This publication is based on observations obtained with the Samuel Oschin Telescope 48-inch and the 60-inch Telescope at the Palomar
Observatory as part of the Zwicky Transient Facility project. ZTF is supported by the National Science Foundation under Grants
No. AST-1440341 and AST-2034437 and a collaboration including current partners Caltech, IPAC, the Oskar Klein Center at
Stockholm University, the University of Maryland, University of California, Berkeley , the University of Wisconsin at Milwaukee,
University of Warwick, Ruhr University, Cornell University, Northwestern University and Drexel University. Operations are
conducted by COO, IPAC, and UW.

Guoshoujing Telescope (the Large Sky Area Multi-Object Fiber Spectroscopic Telescope LAMOST) is a National Major Scientific Project built by the Chinese Academy of Sciences. Funding for the project has been provided by the National Development and Reform Commission. LAMOST is operated and managed by the National Astronomical Observatories, Chinese Academy of Sciences.

\clearpage

\appendix
\counterwithin{figure}{section}
\counterwithin{table}{section}

\section{Figures}

\begin{figure}[!hbtp]
    \centering
    \includegraphics[width=1.0\linewidth]{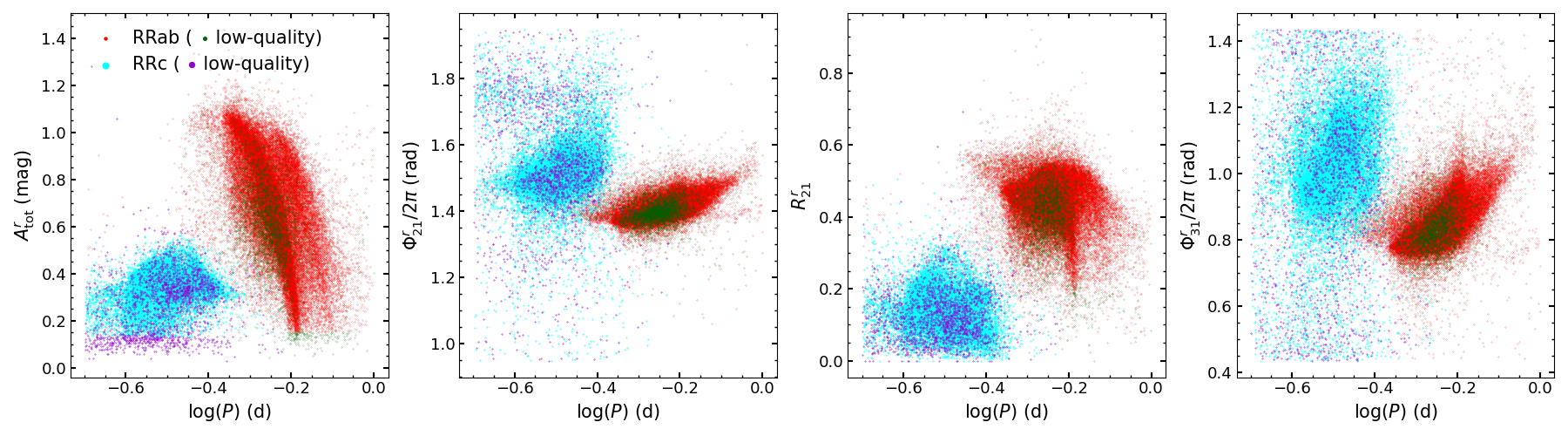}
    \caption{Similar to Figure\,\ref{fig:Fourier_g}, but showing the Fourier parameters with periods for the ZTF\_RRL\_ALL sample in the $r$-band.}
\label{fig:Fourier_r}
\end{figure}

\begin{figure}[!hbtp]
    \centering
    \includegraphics[width=1.0\linewidth]{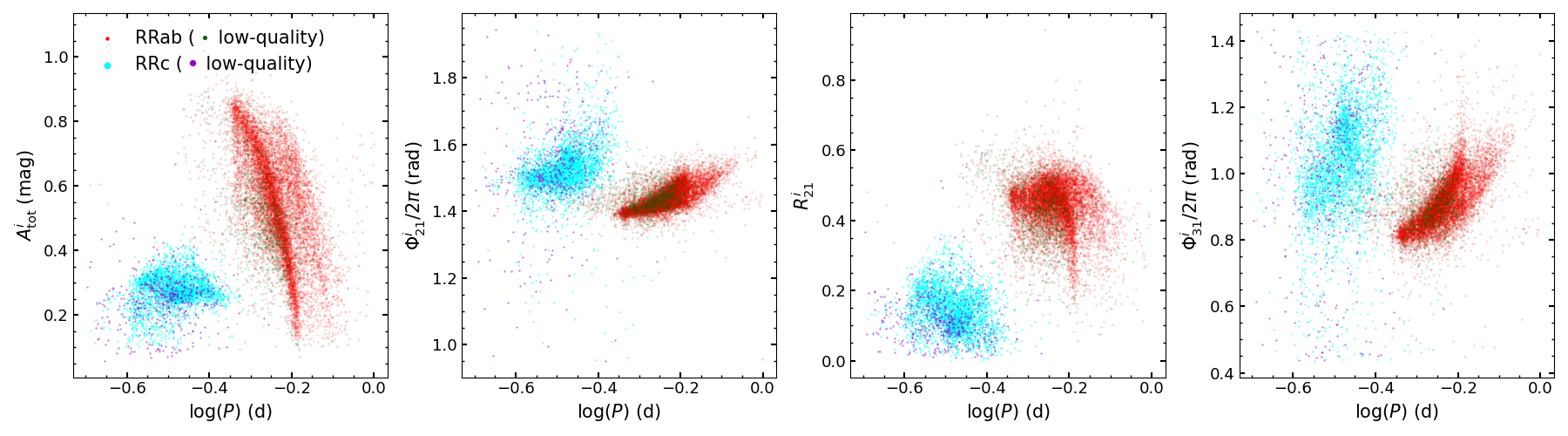}
    \caption{Similar to Figure\,\ref{fig:Fourier_g}, but showing the Fourier parameters with periods for the ZTF\_RRL\_ALL sample in the $i$-band.}
\label{fig:Fourier_i}
\end{figure}

\begin{figure*}[!htbp]
\centering
\includegraphics[width=0.98\linewidth]{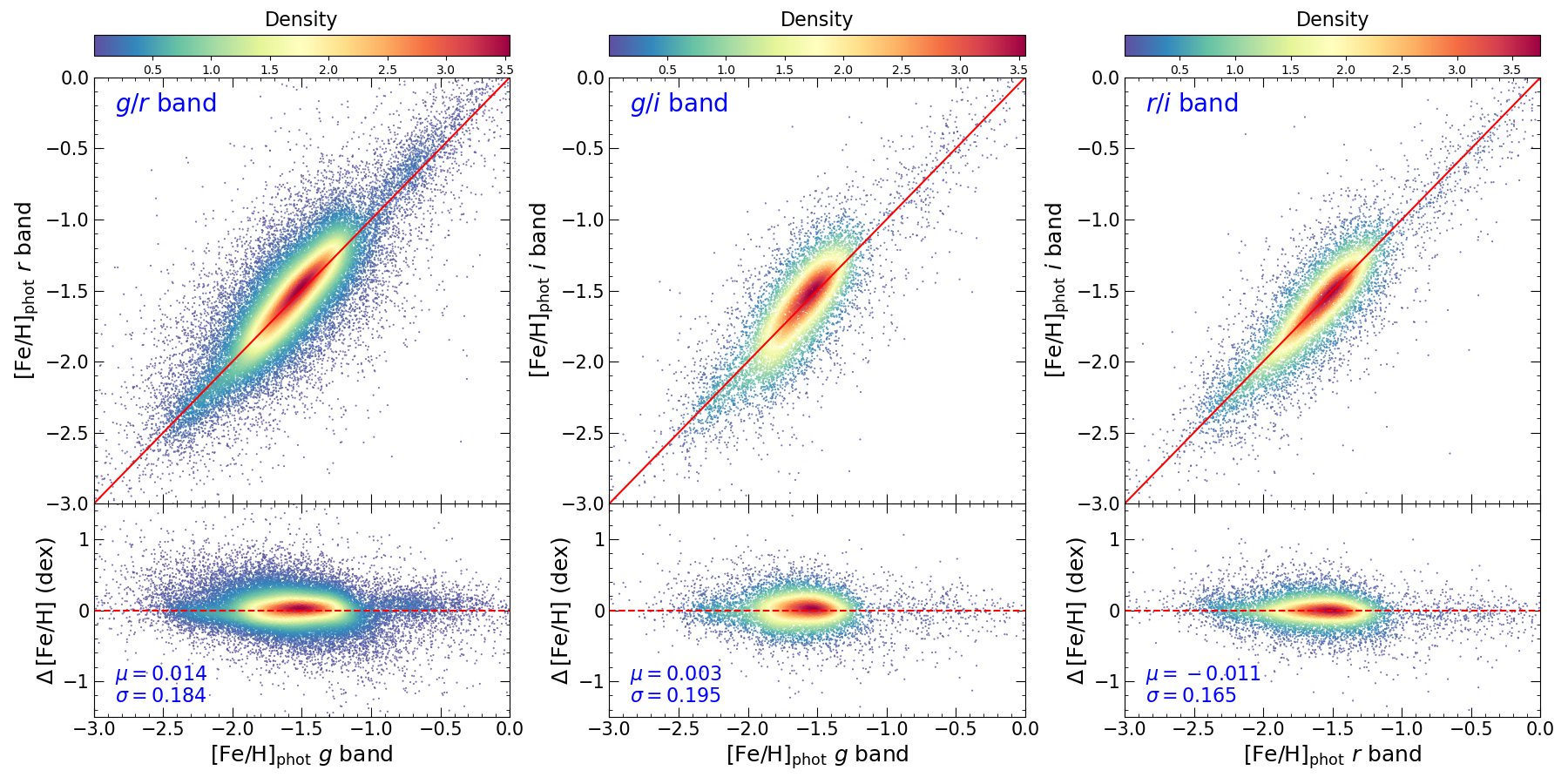}
\caption{Upper section of panels: Photometric-metallicity estimates from this work for RRab stars compared between two different bands, labeled the top left. Data are presented as density maps after excluding sample stars with large errors ($>$ 0.5\,dex) to clearly show the overall consistency. The solid-red line is the one-to-one line. Lower section of panels: $\Delta$[Fe/H] differences (y-axis minus x-axis), with the mean and standard deviation shown in the bottom left corner of each panel.  The dashed-red line is the zero residual level.}
\label{fig:metal_comself_ab}
\end{figure*}

\begin{figure*}[!htbp]
\centering
\includegraphics[width=0.98\linewidth]{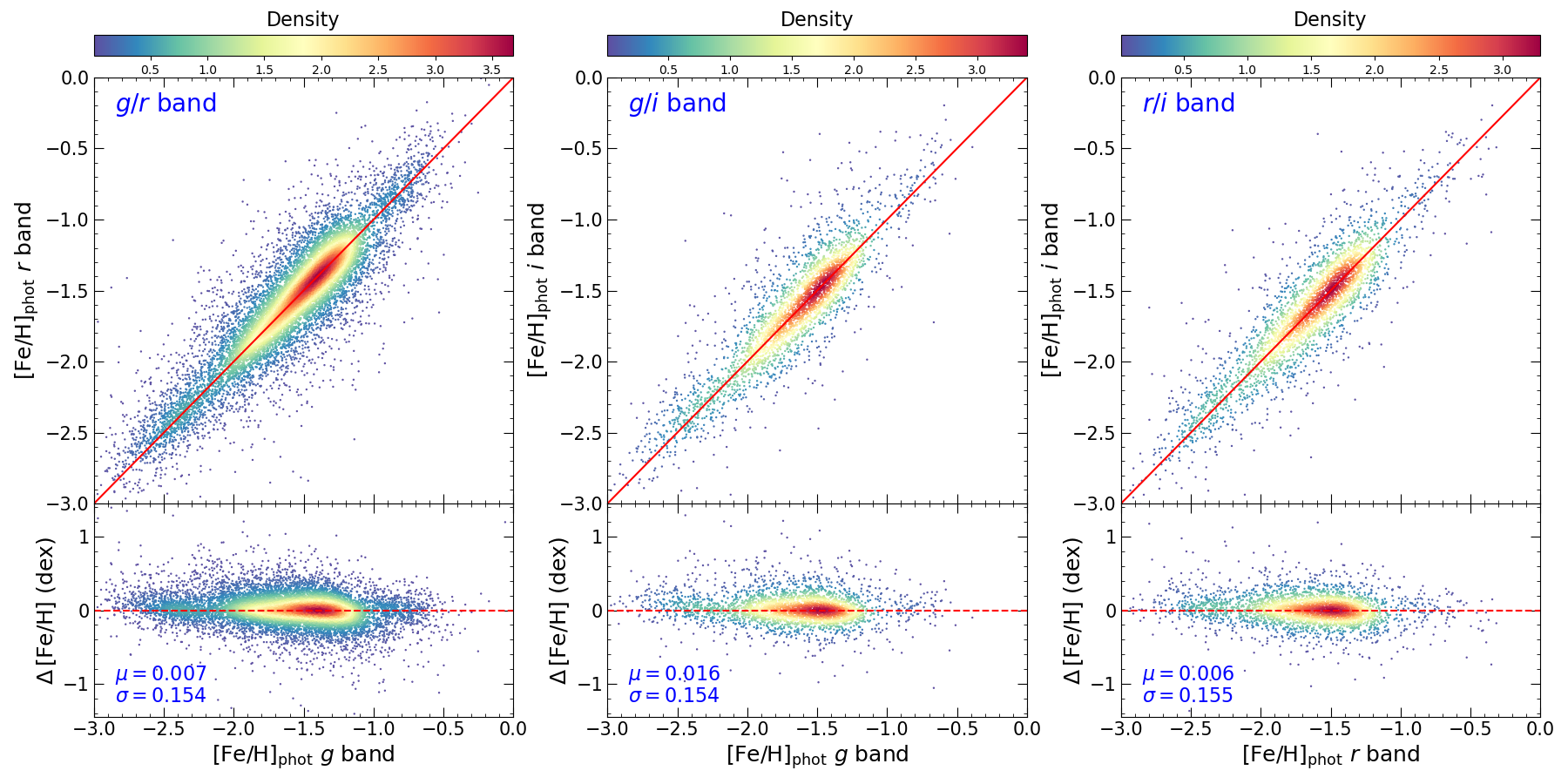}
\caption{Similar to Figure\,\ref{fig:metal_comself_ab}, but for type RRc stars.}
\label{fig:metal_comself_c}
\end{figure*}

\begin{figure*}[!htbp]
\centering
\includegraphics[width=0.98\linewidth]{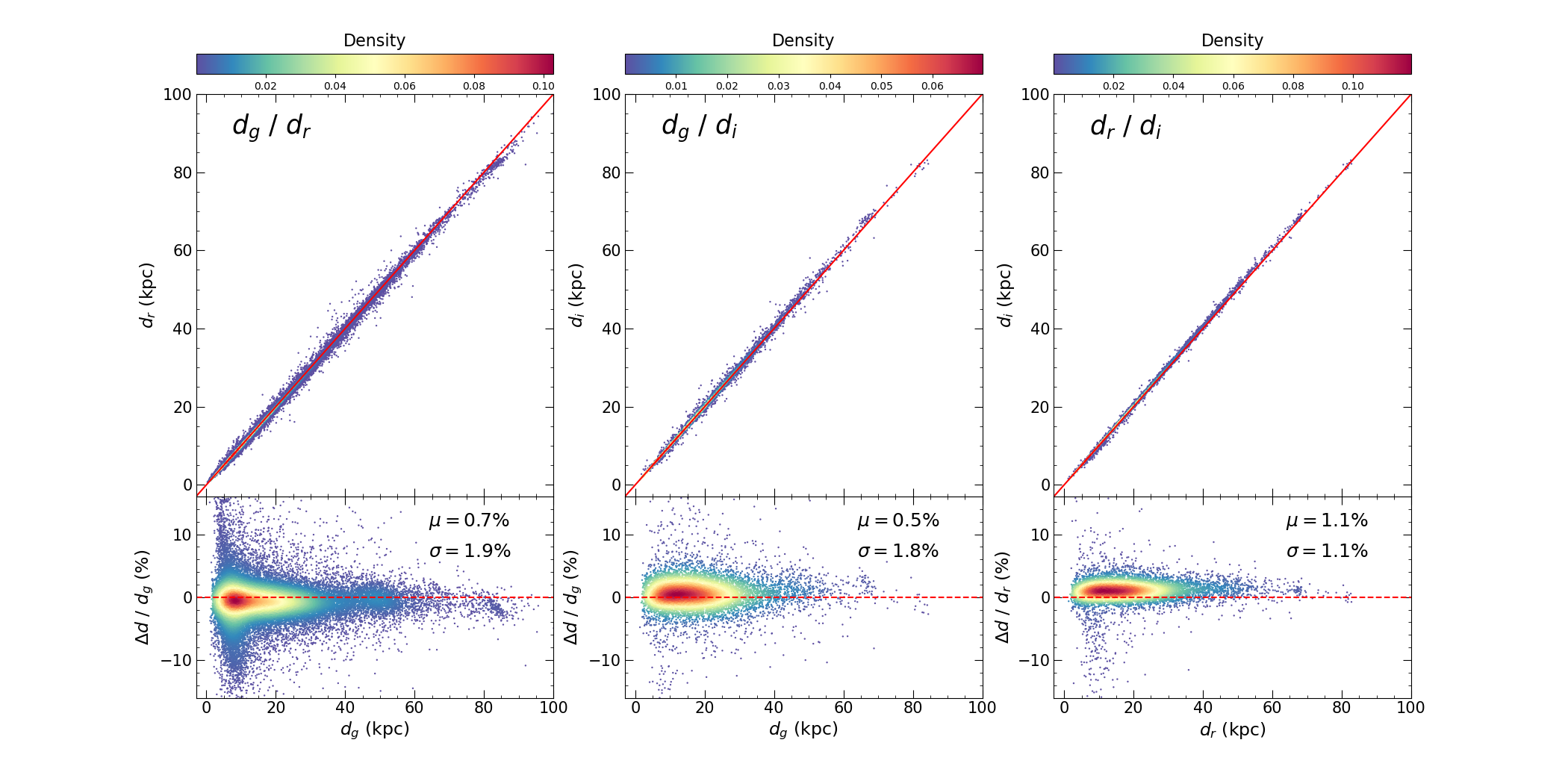}
\caption{Upper section of panels: Distances derived from the PMZ relations in this study are compared to a density map of RRab stars across two different bands, as labeled in the top left. The increased local dispersion shown in the figure is primarily attributed to regions with high extinction, where the use of A$_G$ results in larger extinction errors, affecting the accuracy of our distance estimates. The solid-red line is the one-to-one line. 
Lower section of panels: The relative difference of the y-axis distance with respect to the x-axis distance, with the mean and standard deviation of these relative differences shown in the upper right corner of each panel. The dashed-red line is the zero level.}
\label{fig:pmz_dist_comself_ab}
\end{figure*}

\begin{figure*}[!htbp]
\centering
\includegraphics[width=0.98\linewidth]{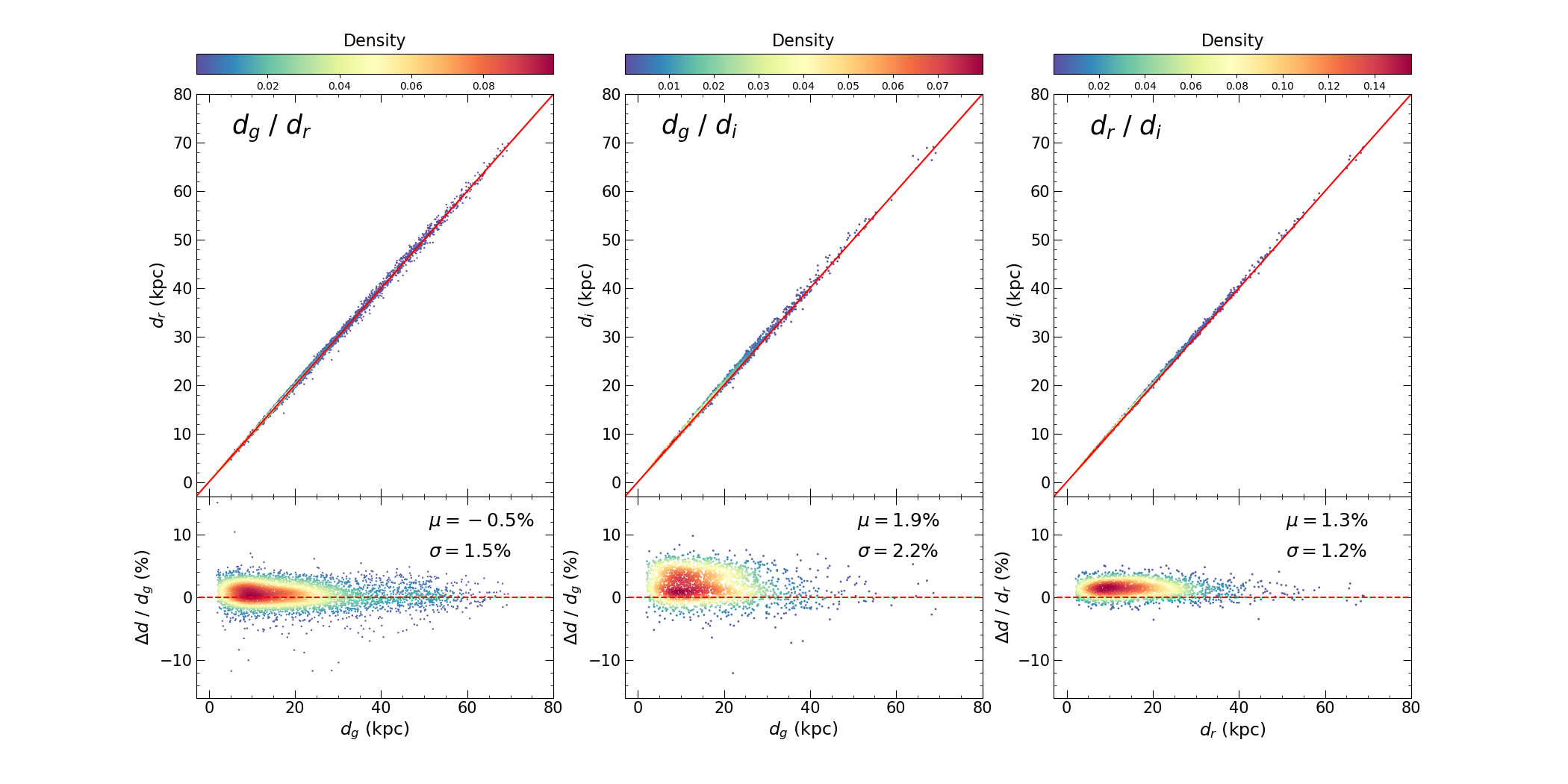}
\caption{Similar to Figure\,\ref{fig:pmz_dist_comself_ab}, but for type RRc stars. Note that the extinction values are sourced solely from the SFD98 map, as C23 does not provide absorption data for RRc stars.}
\label{fig:pmz_dist_comself_c}
\end{figure*}

\begin{figure*}[!htbp]
\centering
\includegraphics[width=0.98\linewidth]{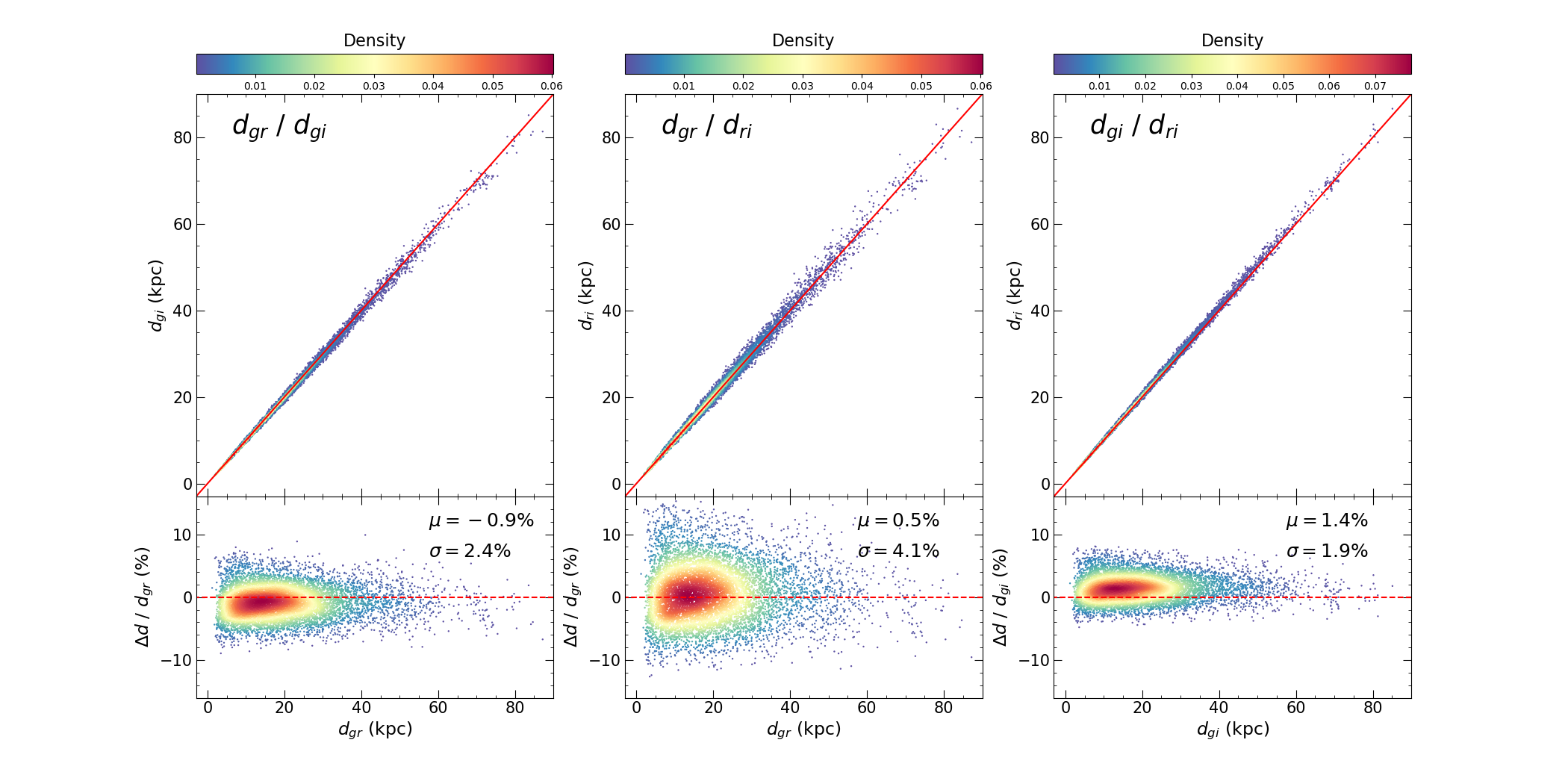}
\caption{Upper section of panels: Distances derived from the PWZ relations in this study are compared to a density map of RRab stars across two different bands, as labeled in the top left. The solid-red line is the one-to-one line.  Lower section of panels: The relative difference of the y-axis distance with respect to the x-axis distance, along with the mean and standard deviation of these relative differences displayed in the bottom right corner of each panel. The dashed-red line is the zero level.}
\label{fig:pwz_dist_comself_ab}
\end{figure*}

\begin{figure*}[!htbp]
\centering
\includegraphics[width=0.98\linewidth]{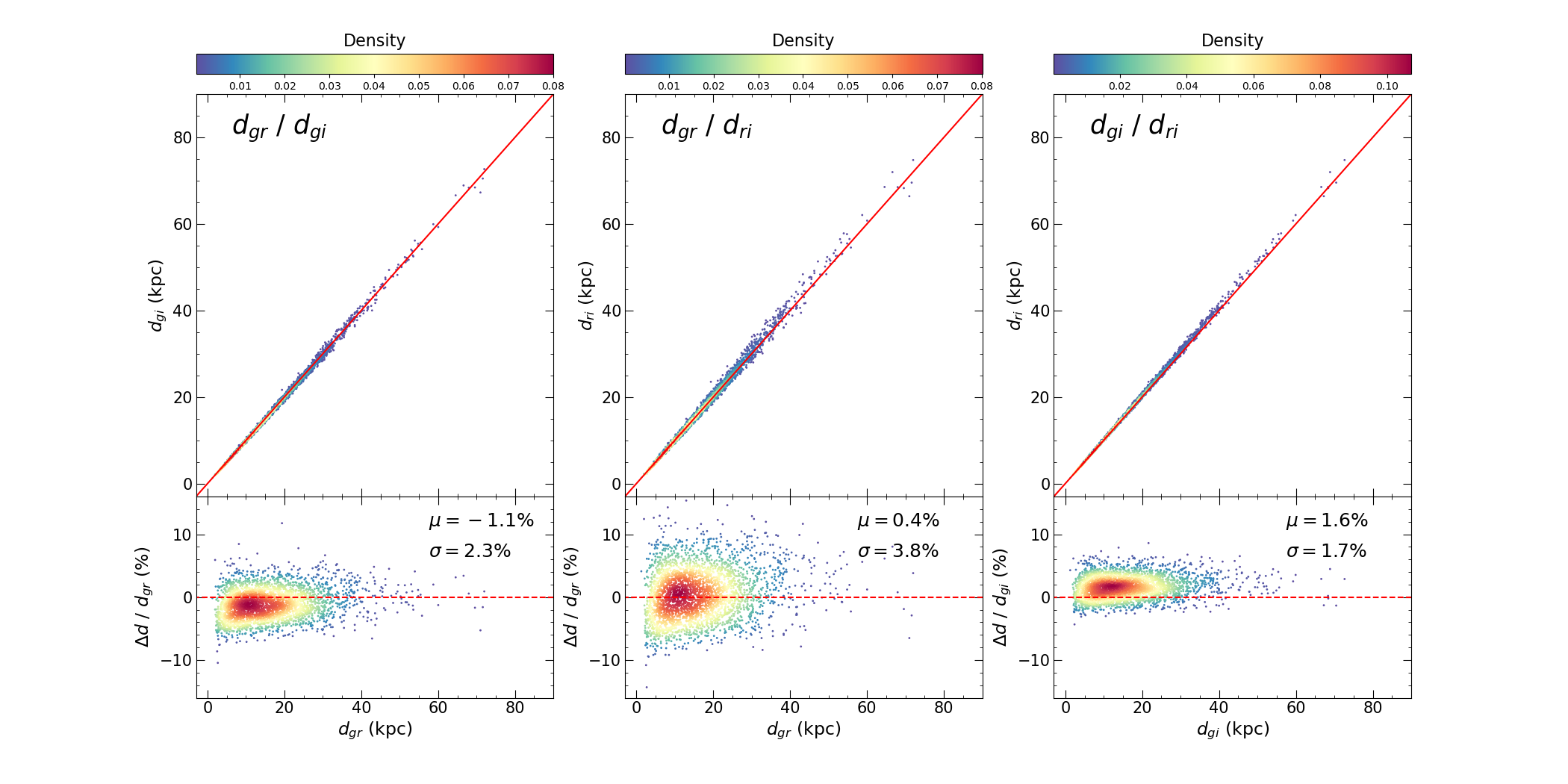}
\caption{Similar to Figure\,\ref{fig:pwz_dist_comself_ab}, but for type RRc stars.}
\label{fig:pwz_dist_comself_c}
\end{figure*}

\begin{figure*}[!htbp]
\centering
\includegraphics[width=0.98\linewidth]{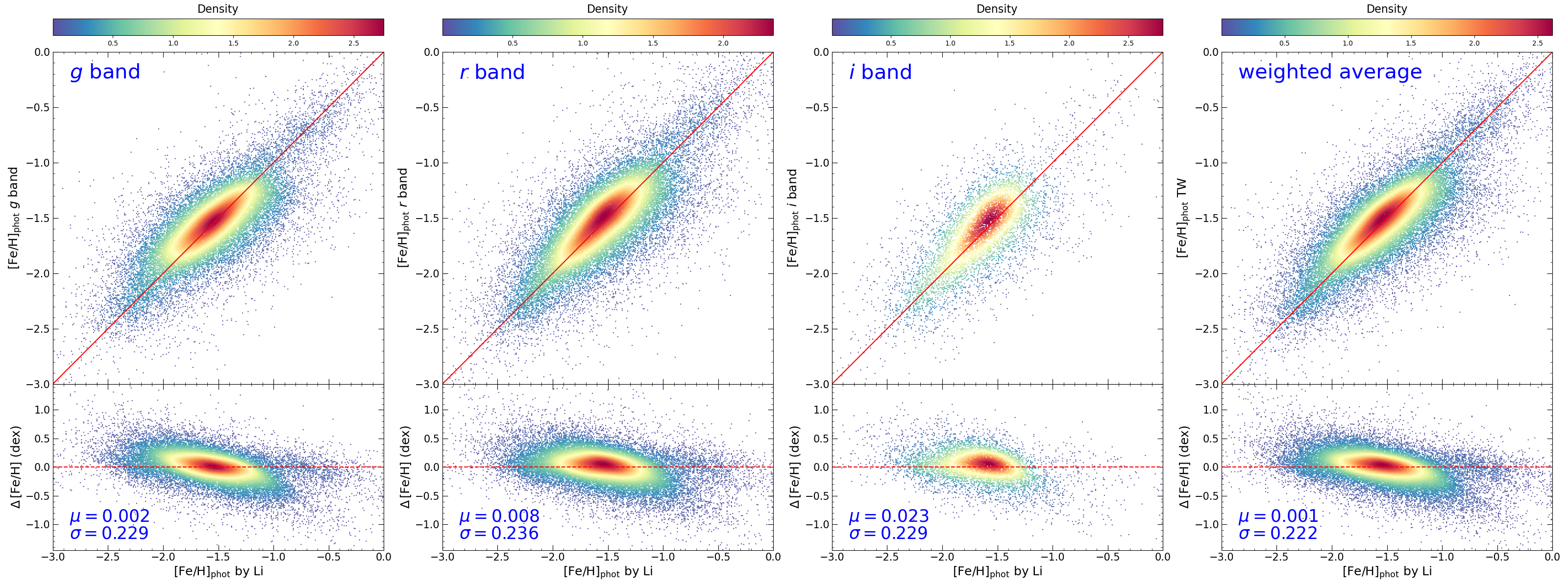}
\caption{Upper section of panels: Photometric-metallicity estimates for RRab stars compared between the Li et al. work and this study, with the $gri$-bands, and weighted average of the three bands displayed from left to right. The data are presented as a density map after excluding sample stars with large errors ($>$ 0.5\,dex) for clearly showing the overall consistency. The solid-red line is the one-to-one line.  Lower section of panels: $\Delta$[Fe/H] differences (our work minus Li et al.), with the mean and standard deviation shown in the in the bottom left corner of each panel.  The dashed-red line is the zero level.}
\label{metal_cpr_li_ab}
\end{figure*}

\begin{figure*}[!htbp]
\centering
\includegraphics[width=0.98\linewidth]{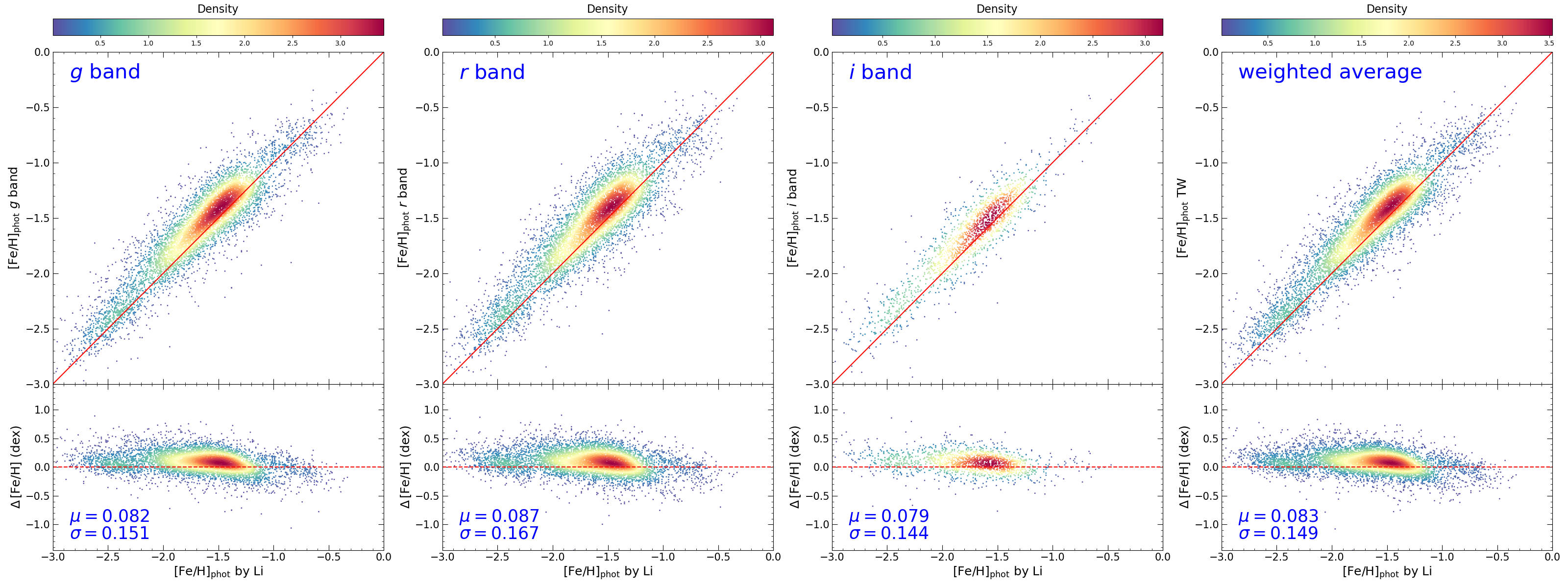}
\caption{Similar to Figure\,\ref{metal_cpr_li_ab}, but for type RRc stars.}
\label{metal_cpr_li_c}
\end{figure*}

\begin{figure*}[!htbp]
\centering
\includegraphics[width=0.98\linewidth]{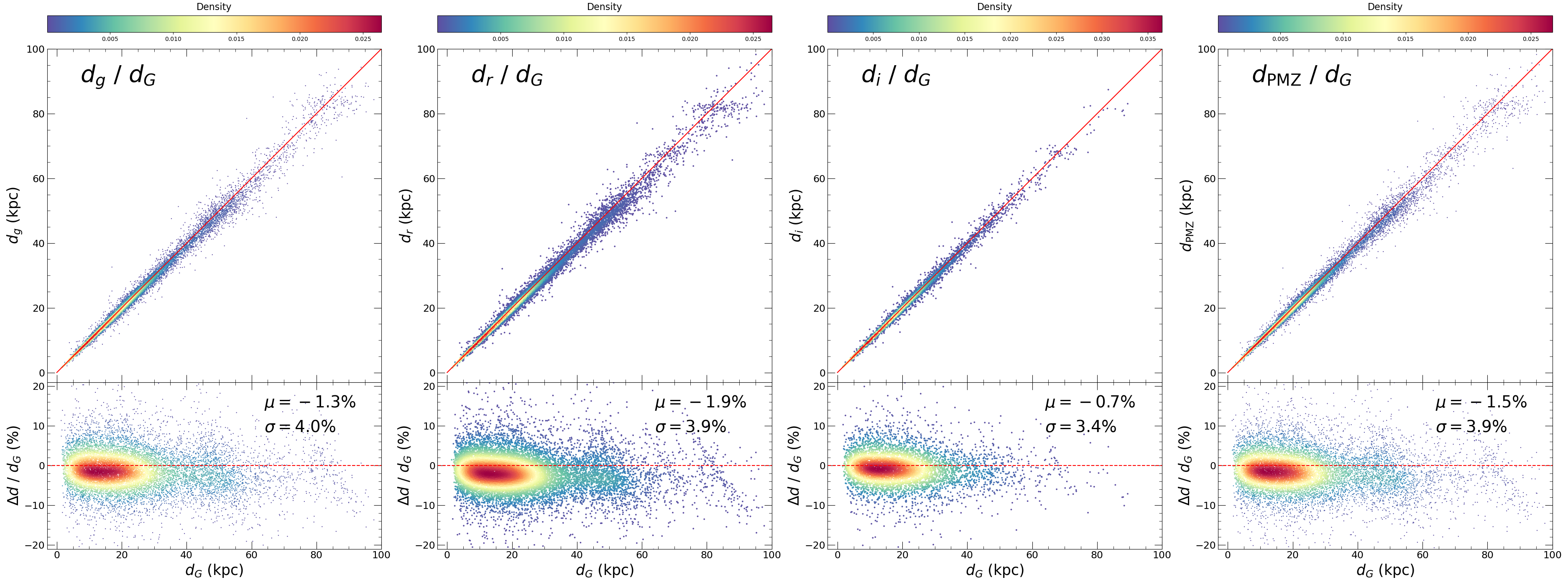}
\caption{Upper section of panels: Distance estimates for RRab stars compared between the Li et al. work and this study using the PMZ relations, with results for the $gri$-bands, and their weighted average displayed from left to right. The solid-red line is the one-to-one line. Lower section of panels: The relative difference of our calculated distance with respect to the referenced distance, along with the mean and standard deviation of these relative differences displayed in the upper right corner of each panel. The dashed-red line is the zero level.}
\label{dist_pmz_li_ab}
\end{figure*}

\begin{figure*}[!htbp]
\centering
\includegraphics[width=0.98\linewidth]{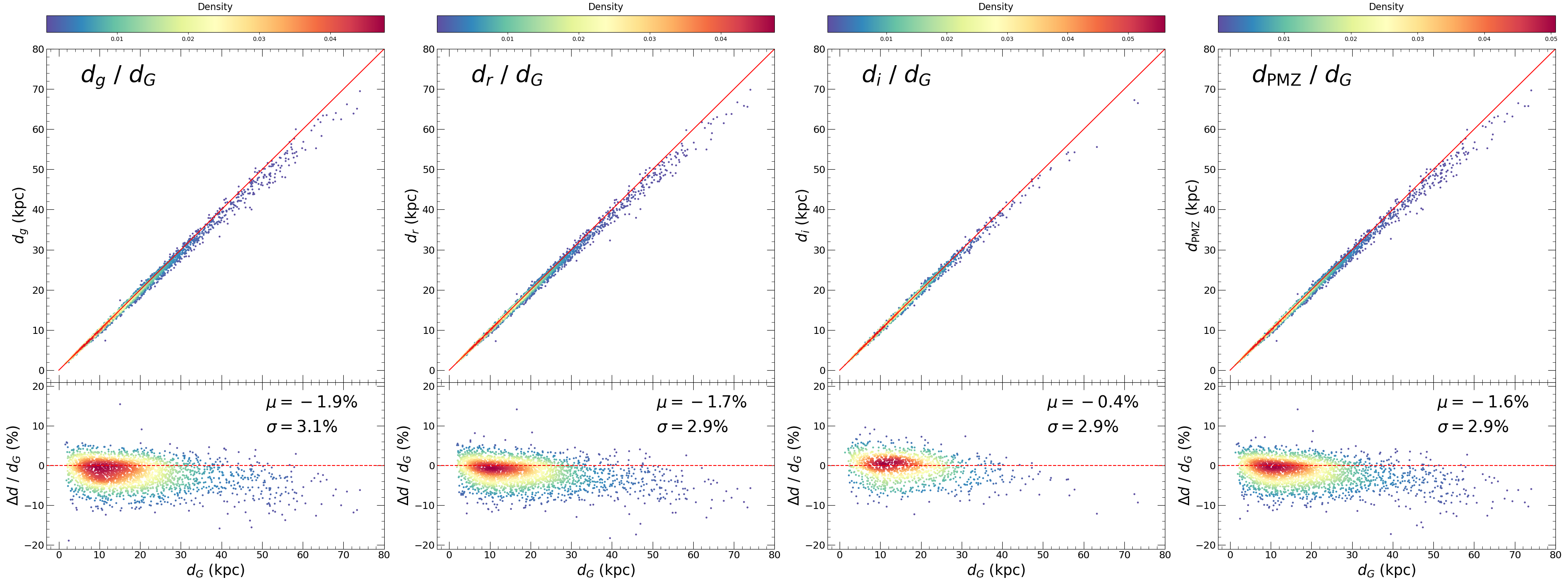}
\caption{Similar to Figure\,\ref{dist_pmz_li_ab}, but for type RRc stars.}
\label{dist_pmz_li_c}
\end{figure*}

\begin{figure*}[!htbp]
\centering
\includegraphics[width=0.98\linewidth]{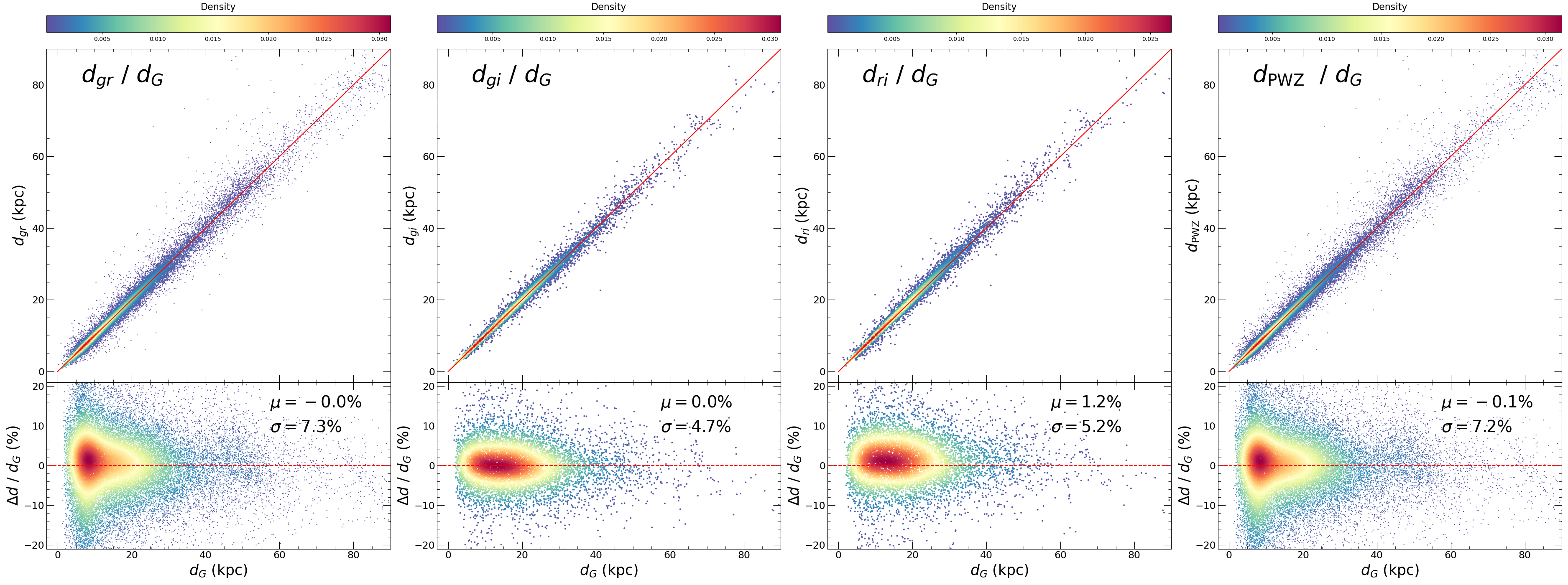}
\caption{Upper section of panels: Comparison of distance estimates for RRab stars between the Li et al. work and our newly constructed PWZ relations, with results for the $d_{gr}$, $d_{gi}$, $d_{ri}$ and their weighted average distance displayed from left to right. The solid-red line is the one-to-one line. Lower section of panels: The relative difference of calculated distance with respect to referenced distance, along with the mean and standard deviation of these relative differences displayed in the upper right corner of each panel. The dashed-red line is the zero level.}
\label{dist_pwz_li_ab}
\end{figure*}

\begin{figure*}[!htbp]
\centering
\includegraphics[width=0.98\linewidth]{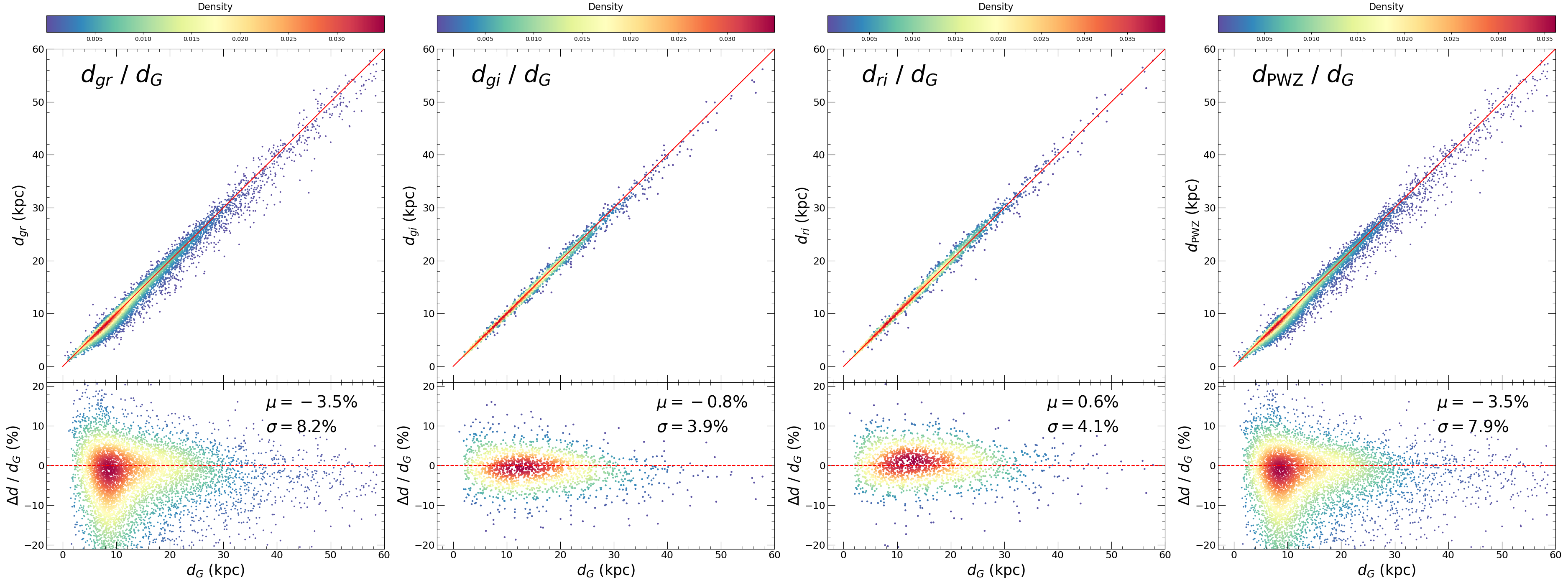}
\caption{Similar to Figure\,\ref{dist_pwz_li_ab}, but for type RRc stars.}
\label{dist_pwz_li_c}
\end{figure*}

\begin{figure}[!hbtp]
    \centering
    \includegraphics[width=1\linewidth]{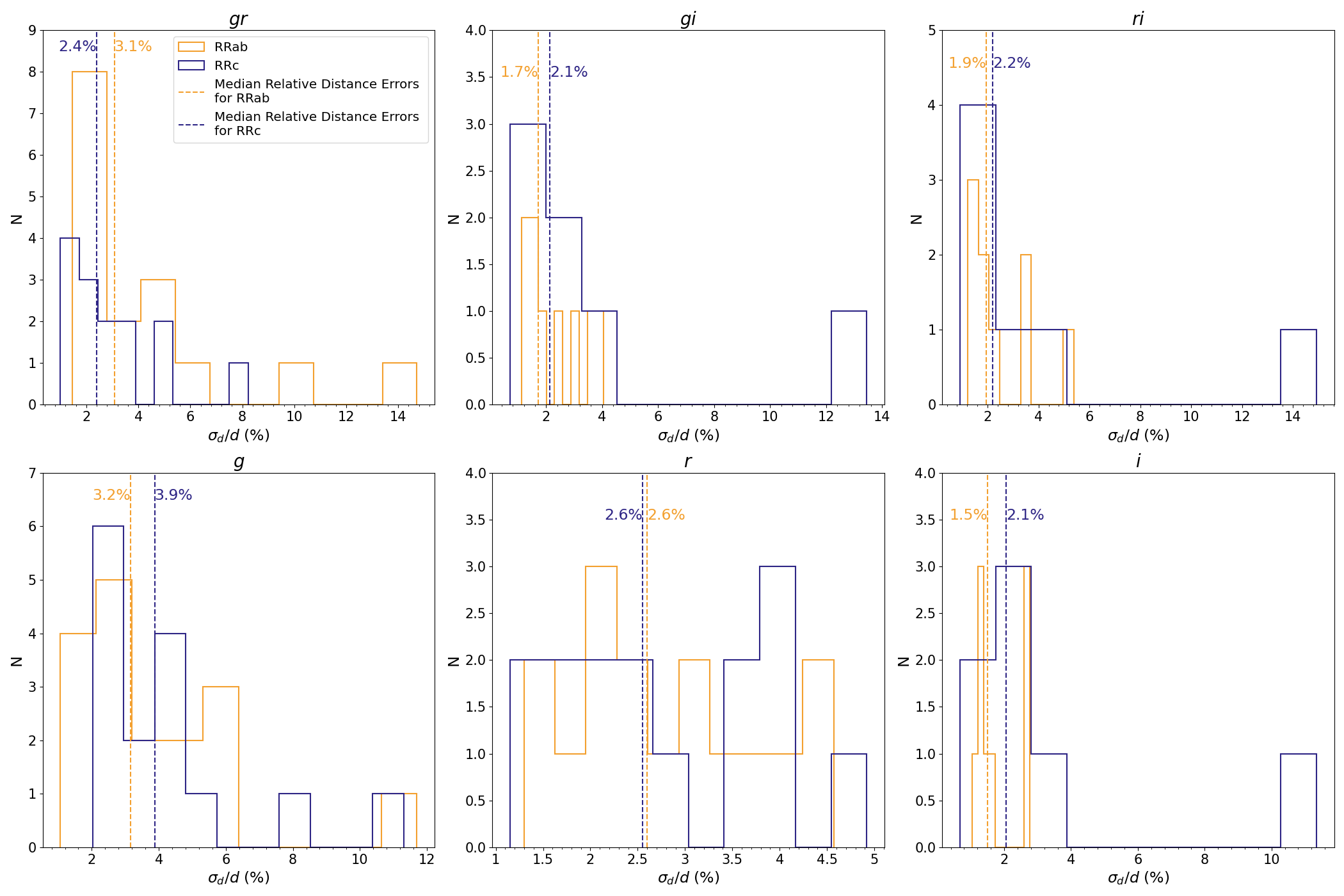}
    \caption{Similar to Figure\,\ref{fig:clu_pmzpwz_sig}, but showing the relative errors of distances for individual $gri$-bands, and for the $gr$-, $gi$-, $ri$-band pairs.}
\label{fig:clu_dist_6sig}
\end{figure}

\begin{figure*}[!htbp]
\centering
\includegraphics[width=0.98\linewidth]{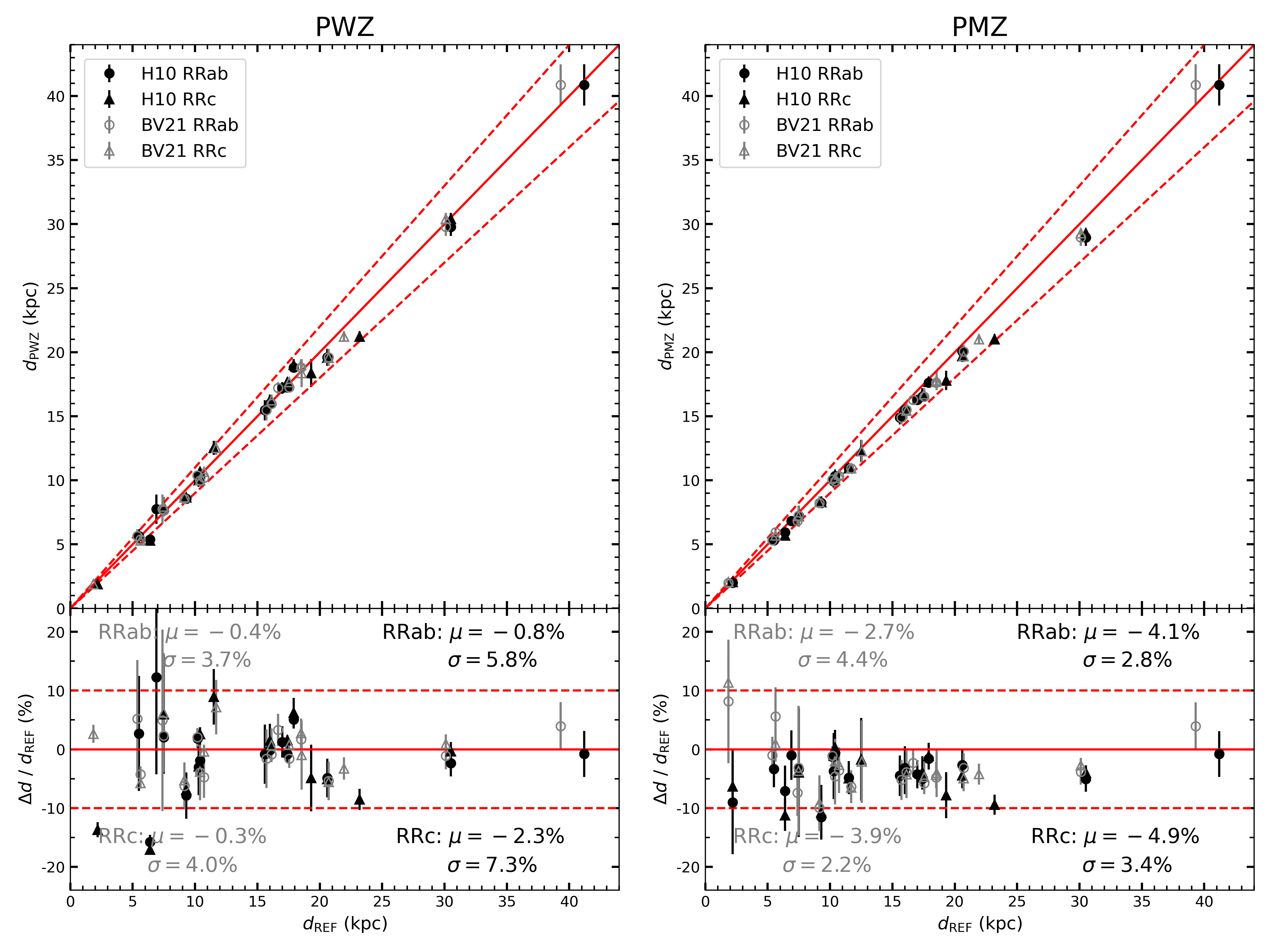}
\caption{Left panel: Comparison of distance estimates from reference works, BV21 (in gray) and H10 (in black), with those derived from our newly constructed PWZ relations, using circles for RRab and triangles for RRc stars. The solid-red line is the one-to-one line, while the two dashed-red lines show 10\% deviations.  The lower section of the panel displays the relative difference of calculated distance with respect to referenced GCs' distance, along with the mean and standard deviation of these relative differences displayed in the corner. The solid-red line is the zero residual line, while the two red-dashed lines indicate 10\% deviations. 
Right panel: A similar comparison, but for the PMZ relation. }
\label{clu_dist_pmzpwz_comp}
\end{figure*}

\begin{figure}[!hbtp]
    \centering
    \includegraphics[width=1.0\linewidth]{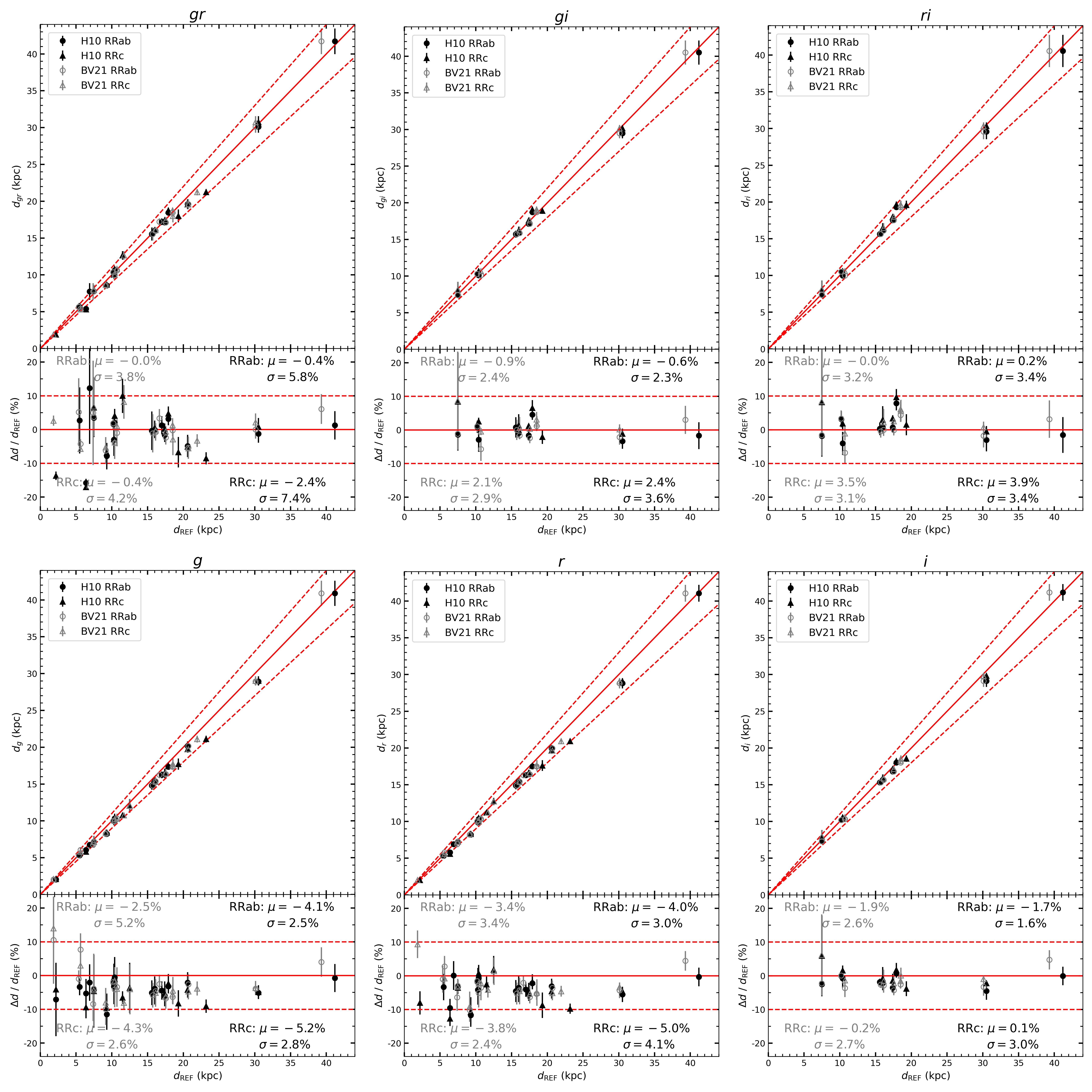}
    \caption{Similar to Figure\,\ref{clu_dist_pmzpwz_comp}, but showing comparisons with reference works for the $gri$-bands, as well as for the $gr$-, $gi$-, $ri$-band pairs.}
\label{fig:clu_dist_6comp}
\end{figure}

\clearpage

\section{Tables}

\begin{table*}[!hbtp]
\tablewidth{\textwidth}
\centering
\caption{Description of the Fourier Parameters Catalog\label{tbl:Fouriercat}}
\renewcommand{\arraystretch}{1.7} 
\begin{threeparttable}
\setlength{\tabcolsep}{22pt} 
\begin{tabular}{lccc}

\hline\hline
&Column&Unit&Description\\

\hline
 \noalign{\smallskip}
  1  &  ID\tnote{a}  &  \dots  & A unique object id for the cataloged RRLs \\
  2  &  SourceID\tnote{b}  &  \dots  & Gaia DR3 source\_id \\
  3  &  RAdeg  &  degree  & Right ascension \\
  4  &  DECdeg  &  degree  & Declination \\
  5  &  GLdeg &  degree  & Galactic longitude \\
  6  &  GBdeg &  degree  & Galactic latitude \\
  7  &  BestP &  d & Average period from light curve in multiple bands\\
  8  &  Nep-gri &  \dots  & Amount of the photometric data actually used in Fourier analysis in $gri$-band\\
  9  &  Totamp-gri &  mag  & $A_{\rm tot}$, total amplitude of the light curve in $gri$-band  \\
  10  &  A1-gri  &  mag  & $A_{1}$, first Fourier amplitudes in $gri$-band \\
  11  &  e\_A1-gri &  mag  &  Uncertainty of the first Fourier amplitudes in $gri$-band \\
  12  &  A2-gri  &  mag  &  $A_2$, second Fourier amplitudes in $gri$-band \\
  13  &  e\_A2-gri  &  mag  & Uncertainty of the second Fourier amplitudes in $gri$-band \\
  14  &  A3-gri &  mag  &   $A_3$, third Fourier amplitudes in $gri$-band \\
  15  &  e\_A3-gri  &  mag  & Uncertainty of the third Fourier amplitudes in $gri$-band \\
  16  &  R21-gri  &  \dots  & Ratio of the $A_2$ and $A_1$ in $gri$-band \\
  17  &  e\_R21-gri &  \dots  & Uncertainty of the ratio of the $A_2$ and $A_1$ in $gri$-band  \\
  18  &  R31-gri &  \dots  & Ratio of the $A_3$ and $A_1$ in $gri$-band \\
  19  &  e\_R31-gri  &  \dots  & Uncertainty of the ratio of the $A_3$ and $A_1$ in $gri$-band   \\
  20  &  Phi1-gri &  rad  & $\phi_1$, first Fourier phases in $gri$-band   \\
  21  &  e\_Phi1-gri &  rad  & Uncertainty of the first Fourier phases in $gri$-band \\
  22  &  Phi2-gri  &  rad  & $\phi_2$, second Fourier phases in $gri$-band \\
  23  &  e\_Phi2-gri  &  rad  & Uncertainty of the second Fourier phases in $gri$-band \\
  24  &  Phi3-gri  &  rad  & $\phi_3$, third Fourier phases in $gri$-band \\
  25  &  e\_Phi3-gri  &  rad  & Uncertainty of the the second Fourier phases in $gri$-band \\
  26  &  Phi21-gri  &  rad  & Phase difference between $\phi_2$ and $\phi_1$ in $gri$-band\\
  27  &  e\_Phi21-gri  &  rad  & Uncertainty of the phase difference between $\phi_2$ and $\phi_1$ in $gri$-band \\
  28  &  Phi31-gri  &  rad  & Phase difference between $\phi_{3}$ and $\phi_{1}$ in $gri$-band \\
  29  &  e\_Phi31-gri  &  rad  & Uncertainty of the phase difference between $\phi_3$ and $\phi_1$ in $gri$-band \\
  30  &  Meanmag-gri  &  mag  & m$_{0}$, mean magnitude in $gri$-band \\
  31  &  e\_Meanmag-gri  &  mag  & Uncertainty of the mean magnitude in $gri$-band \\
  32  &  Phcov-gri  &  \dots  & Phase coverage in $gri$-band phase-folded light curves \\
  33  &  SNR-gri  &  \dots  &  Fit quality indicator in $gri$-band\tnote{c}\\
  34  &  Rstdev-gri  &  mag  & The standard deviation of residuals for fitting light curves in $gri$-band\\
  35  &  Flag\tnote{d}  &  \dots  & Light curve flags with `A', `C', `D'\\
  36  &  Type  &  \dots  & Type of RR Lyrae star, `RRab' or `RRc' \\

\hline
\end{tabular}
\begin{tablenotes}
\item[a] These ids are from the original catalogs.
\item[b] Within a 3 arcsecond radius, only a few sample stars from the Gaia DR3 main catalog did not find matches; these unmatched sample stars are set to 0.
\item[c] Calculated as $\rm Totamp\text{-}g * \sqrt{\rm Nep\text{-}g} / \rm Rstdev\text{-}g$
\item[d] Caution should be exercised when using metallicity and distance parameters for potential low-quality sample stars. It is recommended to verify the light curves in different bands before relying on these values.

(The full table is available in its entirety in machine-readable form.)
\end{tablenotes}
\end{threeparttable}
\end{table*}

\begin{table*}[!hbtp]
\tablewidth{\textwidth}
\centering
\caption{Description of the Final Catalog\label{tbl:fincat}}
\renewcommand{\arraystretch}{1.8} 
\begin{threeparttable}
\begin{tabular}{lccc}

\hline\hline
&Column&Unit&Description\\

\hline
 \noalign{\smallskip}
  1  &  ID\tnote{a}  &  \dots  & A unique object id for the cataloged RRLs \\
  2  &  SourceID\tnote{b}  &  \dots  & Gaia DR3 source\_id \\
  3  &  RAdeg  &  degree  & Right ascension \\
  4  &  DECdeg  &  degree  & Declination \\
  5  &  GLdeg &  degree  & Galactic longitude \\
  6  &  GBdeg &  degree  & Galactic latitude \\
  7  &  [Fe/H]-gri  &  \dots  & Photometric-metallicity in $gri$-band\\
  8  &  e\_[Fe/H]-gri &  \dots  & Uncertainty of the photometric-metallicity in $gri$-band \\
  7  &  [Fe/H]  &  \dots  & Weighted average photometric-metallicity estimates from P-Fourier Params-[Fe/H] relations in $gri$-bands\\
  8  &  e\_[Fe/H] &  \dots  & Uncertainty of the weighted average photometric-metallicity \\
  9   &  Ag / Ar / Ai &  mag  & Extinction values in $gri$-band  \\
  10  &  e\_Ag / e\_Ar / e\_Ai  &  mag  & Uncertainty of the extinction values in $gri$-band \\
  11  &  M-gri &  mag  &  Absolute magnitude in $gri$-band  \\
  12  &  e\_M-gri  &  mag  &  Uncertainty of the absolute magnitude in $gri$-band \\
  13  &  Dist-gri  &  kpc  & Distance from PMZ relation in $gri$-band \\
  14  &  e\_Dist-gri &  kpc  & Uncertainty of the distance from PMZ relation in $gri$-band  \\
  15  &  Dist-PMZ  &  kpc  & Weighted average distance from PMZ relation in $gri$-bands \\
  16  &  e\_Dist-PMZ  &  kpc  & Uncertainty of the weighted average distance from PMZ relation in $gri$-bands \\
  17  &  W-gr / W-gi / W-ri &  mag  & Wesenheit absolute magnitude in $gr$, $gi$, and $ri$-band pair  \\
  18  &  e\_W-gr / e\_W-gi / e\_W-ri &  mag  & Uncertainty of wesenheit absolute magnitude in $gr$, $gi$, and $ri$-band pair \\
  19  &  Dist-gr / Dist-gi / Dist-ri  &  kpc  & Distance from PWZ relation in $gr$, $gi$, and $ri$-band pair  \\
  20  &  e\_Dist-gr / e\_Dist-gi / e\_Dist-ri &  kpc  & Uncertainty of the distance from PWZ relation in $gr$, $gi$, and $ri$-band pair  \\
  21  &  Dist-PWZ &  kpc  & Weighted average distance from PWZ relation in $gr$-, $gi$-, and $ri$-band pairs \\
  22  &  e\_Dist-PWZ  &  kpc  & Uncertainty of the weighted average distance from PWZ relation in $gr$-, $gi$-, and $ri$-band pairs \\
  23  &  Flag\tnote{c}  &  \dots  & Light curve flags with `A', `C', `D'\\
  24  &  Type  &  \dots  & Type of the RR Lyrae stars,`RRab' or `RRc'  \\

\hline
\end{tabular}
\begin{tablenotes}
\item[a] These ids are from the original catalogs.
\item[b] Within a 3 arcsecond radius, only a few sample stars from the Gaia DR3 main catalog did not find matches; these unmatched sample stars are set to 0.
\item[c] Caution should be exercised when using metallicity and distance parameters for potential low-quality sample stars. It is recommended to verify the light curves in different bands before relying on these values.

(The full table is available in its entirety in machine-readable form.)
\end{tablenotes}
\end{threeparttable}
\end{table*}

\clearpage

\bibliographystyle{apj}
\bibliography{ref}

\end{document}